\documentclass[12pt,epsf]{article}

\def\balpha{\mbox{\boldmath$\alpha$}}
\def\bbeta{\mbox{\boldmath$\beta$}}

\def\bsigma{\mbox{\boldmath$\sigma$}}
\def\bgamma{\mbox{\boldmath$\gamma$}}

\def\bp{{\bf p}}

\def\bx{{\bf x}}
\def\by{{\bf y}}

\def\bP{{\bf P}}
\def\bX{{\bf X}}

\def\ba{{\bf a}}
\def\bb{{\bf b}}
\def\bc{{\bf c}}
\newcommand{\be}{\begin{equation}}
\newcommand{\ee}{\end{equation}}
\newcommand{\bea}{\begin{eqnarray}}
\newcommand{\eea}{\end{eqnarray}}

\newcommand{\lan}{\langle}
\newcommand{\ran}{\rangle}

\usepackage[dvips]{graphics}
 
\begin{document}
 
\vskip 2cm
\begin{center}
\Large
{\bf  A Question of } \\
{\bf Self-consistent Semifactuality } \\
\vskip 0.5cm
\large
Ken Williams \\
{\small \sl Department of Physics } \\
{\small \sl University of Wisconsin - Parkside } \\
\end{center}
\thispagestyle{empty}
\vskip 0.7cm

\begin{abstract} 
 
This article is intended as a compendium and guide to the variety of Bell Inequality derivations that have appeared in the literature in recent years, classifying them into six broad categories, revealing the underlying, often hidden, assumption common to each - semifactuality. Evaluation of the attendant conditional brings to light a significant EPR loophole that has not appeared in the literature. Semantics for the inequality in the ongoing philosophic debate that led to its discovery is discussed.

\end{abstract}
 
\newpage

\section{Overview}

There are many derivations of the Bell Inequality in the literature, most if not all of which can be classified into one or more of six categories that sometime overlap. Here we show that when it does not make an explicit appearance, semifactual-definiteness is the implicit assumption key to valid derivation in each category. Next, as our main result, we show the conventional Bell Inequality semifactual to be inconsistent, whereupon we present a self-consistent one which on evaluation reveals a highly restrictive constraint necessary for application of the inequality to EPR experimental data. Finally we consider the meaning of the inequality and offer criticism of the philosophic debate that led to its discovery.

In section 1 we motivate and in 2 review the EPR paradox. Sections 3.1 and 3.2 follow with distinct Bell inequality derivations; the next derivation in section 3.3 may be understood as a algebraic variation of either of the preceding two. Section 4 points out the shortcoming common to these derivations and presents in the derivations of 4.1 and 4.2 the random detector-setting condition as a remedy. Implementation of the condition is considered in section 5 by way of an analysis of the commonly associated semifactual. This semifactual is found to be inconsistent with EPR experimental realities, and in 5.1 we introduce in its place an appropriate self-consistent semifactual. In section 6 we consider the resulting experimental constraints and conduct a concluding discussion in section 7. In 7.1 we leave the student with a final thought by way of a popular style of Bell Inequality derivation - there a variant of the elementary set theory derivation of 3.2.

\section{Indeterminacy}

Young's double-slit experiment provides the classic and excellent illustration of the interplay between randomness and correlation peculiar to small-scale phenomena. A stream of identical electrons emitted by a source falls on an opaque screen pierced by two narrow slits and then illuminates a second phosphorous screen, where the electrons collectively produce a pattern of alternating bright and dark fringes - bright where they land, dark where they generally do not.
\begin{center}
\scalebox{0.5}[0.5]{\includegraphics{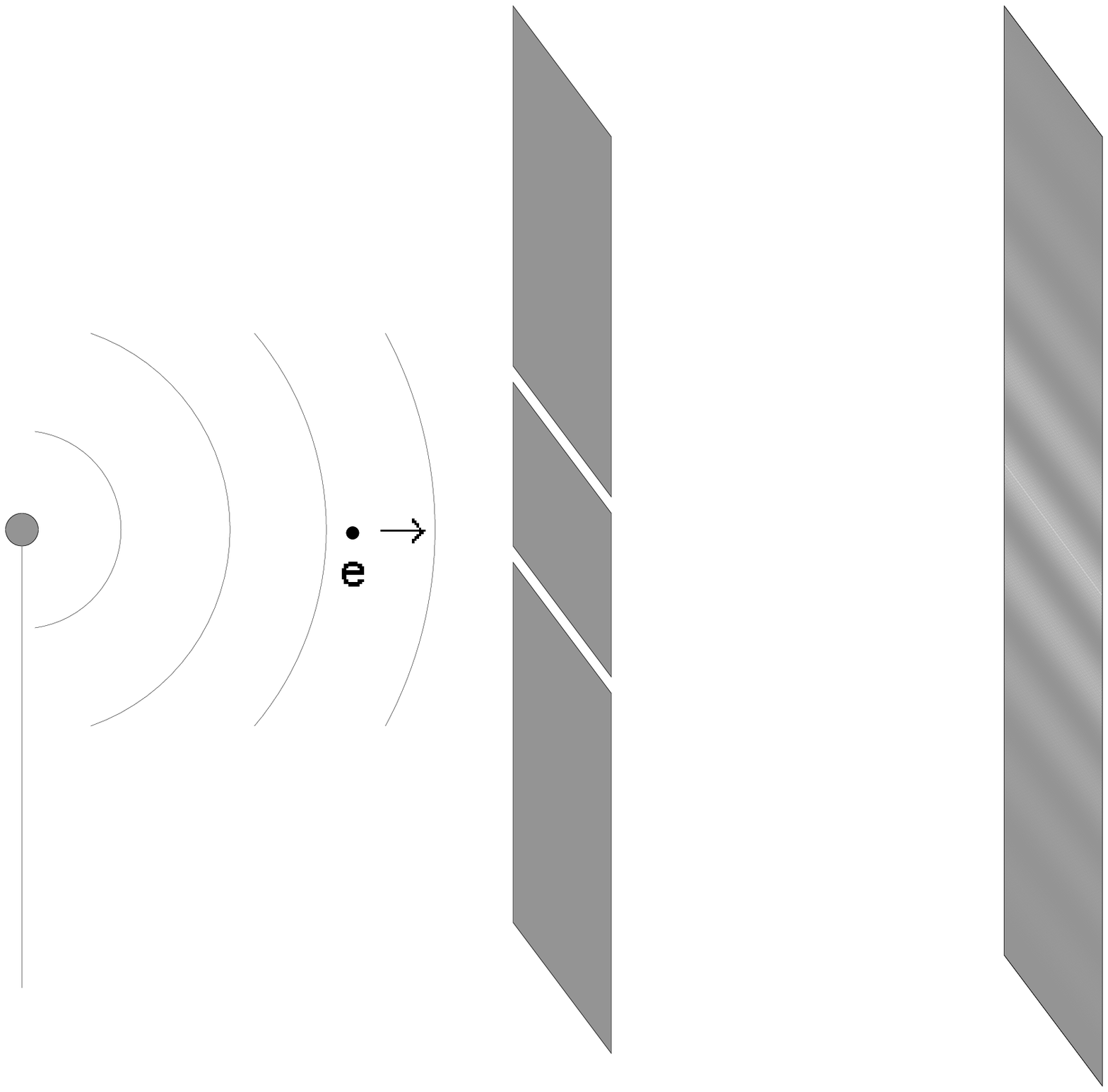}}\\
figure 1
\end{center}
This is the same sort of pattern created by a pair of water-wave point sources (along a surface perpendicular to the bi-section), waves alternatingly in and out of phase, interfering constructively and destructively, respectively
\begin{center}
\scalebox{0.6}[0.6]{\includegraphics{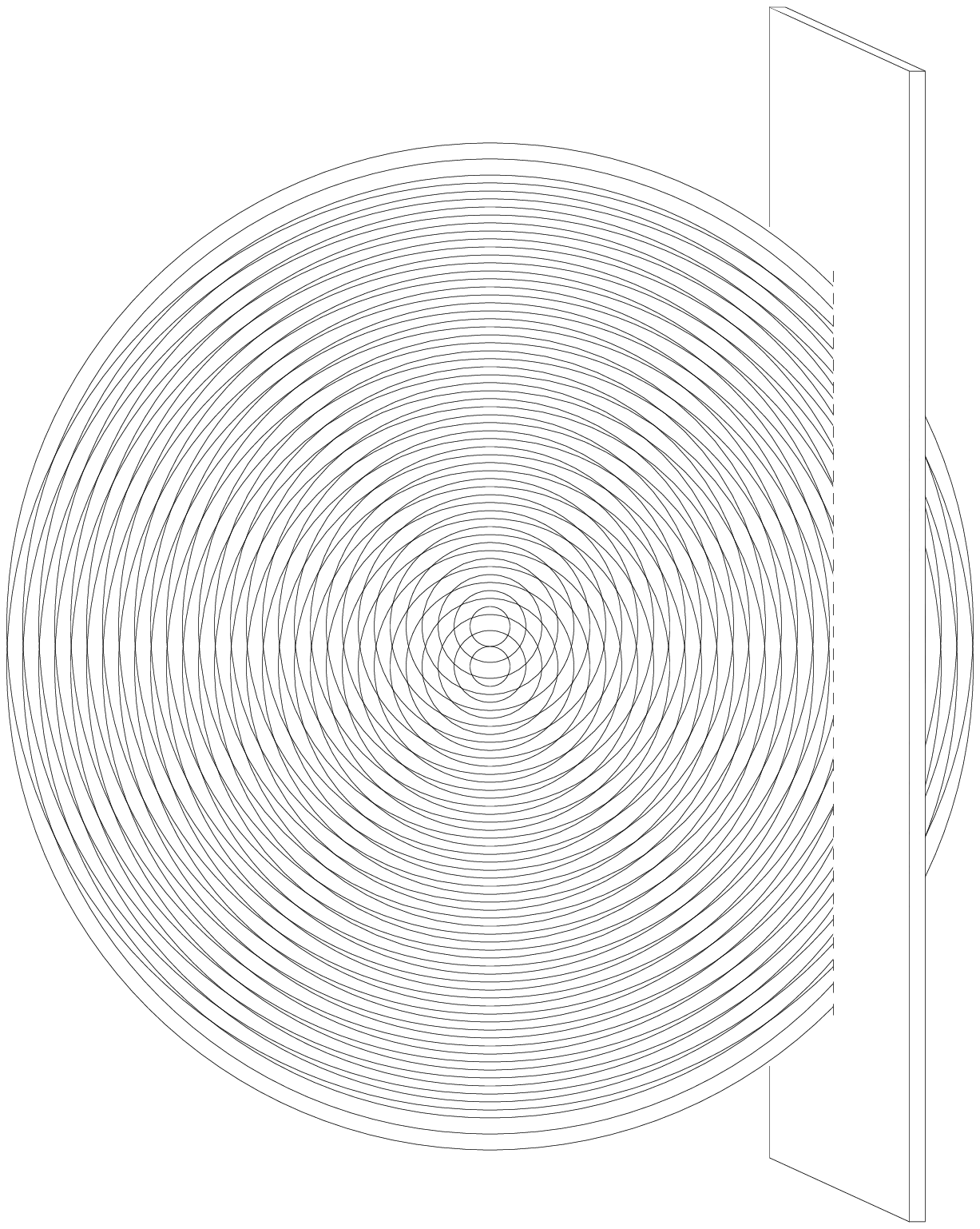}}\\
figure 2
\end{center}
Monochromatic light will also produce this pattern when placed behind the double-slit. 

\noindent The experimental results are
\begin{itemize}
\item When only one slit is open the interference pattern disappears.
\item When the two single slit intensities are superimposed, the interference pattern still does not appear.
\end{itemize}
These observations suggest that the interference results from a phenomena existing between the slits, perhaps interactions between electrons that pass through opposite slits simultaneously. To test the idea we might send the electrons through at a slow rate, one at a time, separating them by large distances so as to prevent the possibility of such an interaction. When this is done the same interference pattern gradually, mysteriously, emerges with the arrival of electrons (fig 3).
\begin{center}
\scalebox{1.0}[1.0]{\includegraphics{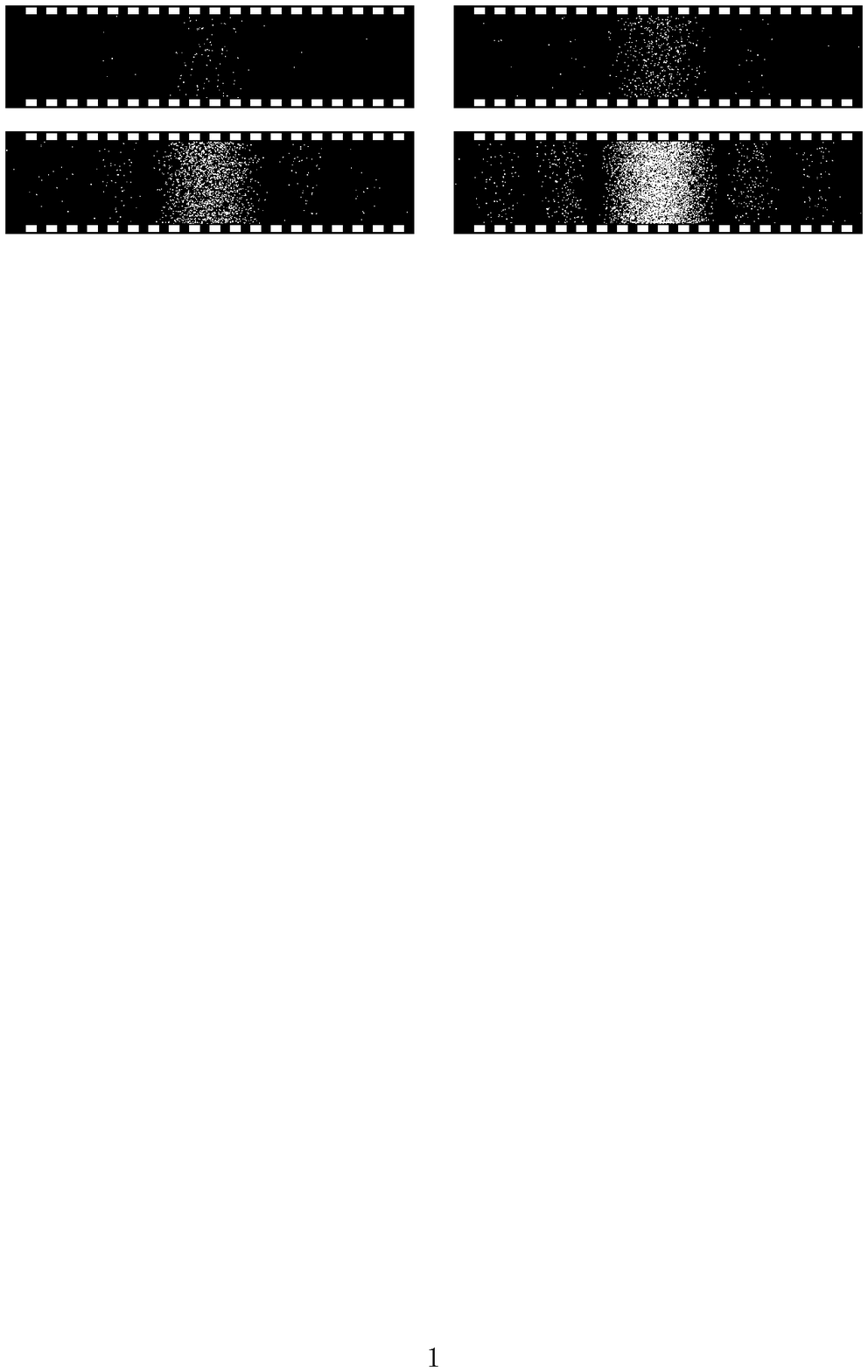}} \\
figure 3
\end{center}
And so it seems that in some sense a single electron passes through both slits simultaneously, interfering with itself much like a wave would. Such an electron has of course no definite trajectory, and the associated spatial uncertainty is manifest at the observation screen where the electron lands unpredictably, though, as also predicted by quantum mechanics, most probably on or near a bright fringe.

\noindent According to the orthodox interpretation of quantum mechanics this unpredictability, or randomness, is fundamental to the measurement process and cannot be gotten around. It does not depend for example on the resolving power of the measuring devices used to take data, but is always present whenever "complementary" quantities are observed, such as an electron's position and momentum. Thus the effective vertical position measurement at the double slits creates a complementary uncertainty in the vertical momentum which from a given slit determines the electron's final position at the observation screen.

\noindent One will often have had occasion to observe similar random behavior in large-scale, or macroscopic, phenomena. Take for example a set of approach-shot lies from a typical practice session of golf. A player on a good run has found his or her touch. Yet, after a series of shots from one spot the lies are never identical, varying randomly over some area 
\begin{center}
\scalebox{0.3}[0.3]{\includegraphics{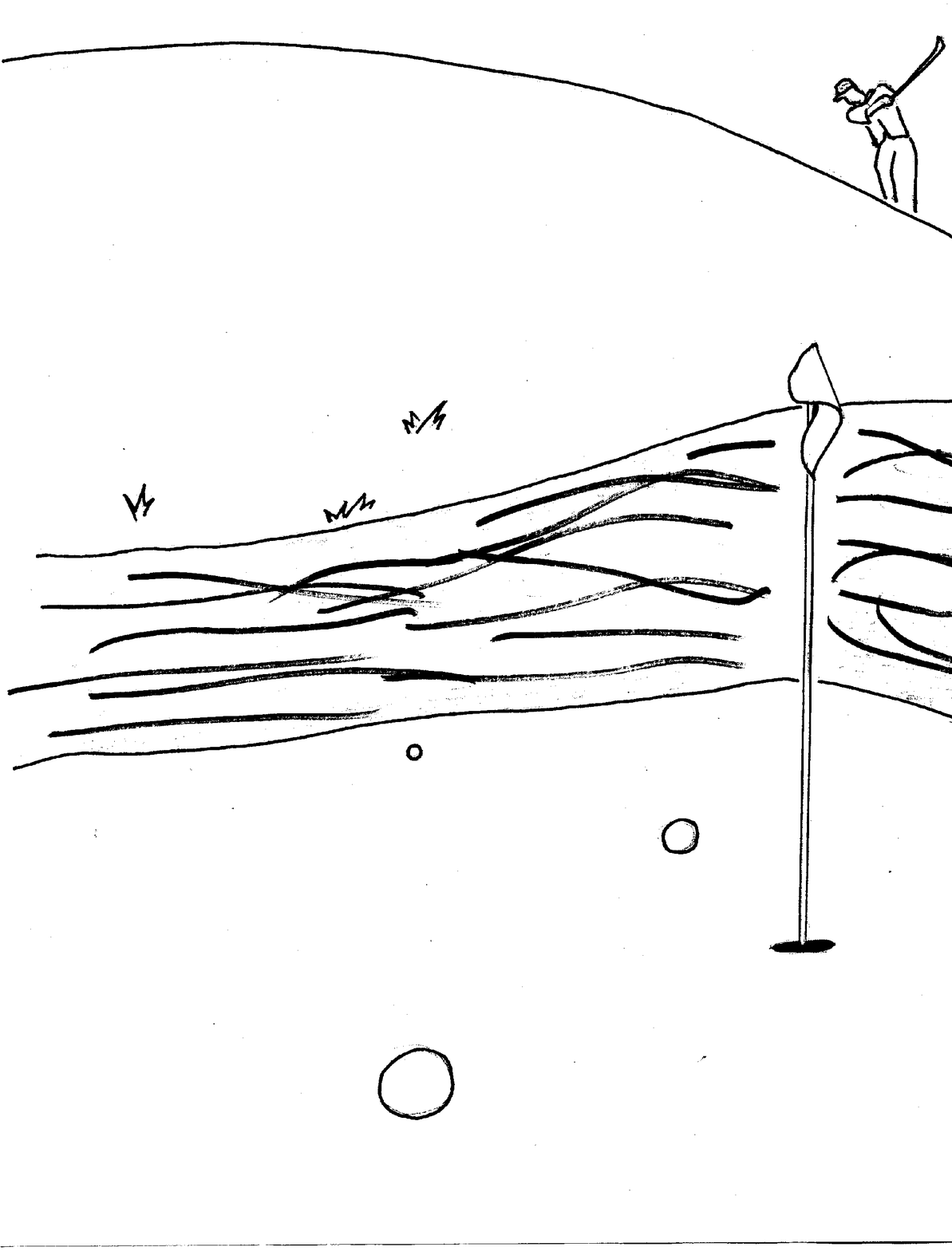}}\\
figure 4
\end{center}
Why? Because of variations in the player's stroke? Maybe. But would there not remain some random variation even if the strokes were identical, say, if the player were replaced by a mechanical swinging device? Probably. Such variation might be due e.g. to wind fluctuation, variation in green speeds, various random vibrations from distant events, etc., any number of less than obvious causes. But what if the mechanical device were programmed to take these factors into account as well so as to progressively refine its stroke? Then one would expect the accuracy also to progressively improve, in principle without limit, lies approaching an ace.

\noindent Physicists are in general agreement that predictions of macroscopic events like these golf shots may be improved upon indefinitely by progressively taking into account the relevant physical variables: wind resistance, orbital motion of the earth, motion of distant galaxies, etc., ad-infinitum. Why then would this not also be true of the microscopic, such as those events in the Young's double-slit experiment? On this question physicists are not in agreement and have not been since the inception of quantum mechanics almost a century ago. Many are not prepared to accept as fundamental the randomness observed in microscopic phenomena, though fewer today than in former times. 

\noindent To these such an acceptance goes against the very program of physics which is to understand the events that make up physical reality - and from understanding, to predict. In this way of thinking then there must exist physically deterministic variables relevant to Young's double-slit experiment and the like, no less than there are for golf, even if these variables are at the moment unknown or hidden from us. Quantum mechanics is thus understood to give an incomplete account of physical reality, just as a theory that described the physics of golf which, however, ignored air resistance also paints a physically incomplete picture, although such a theory might well yield reliable probability predictions.

\section{EPR argument}

Foremost among the dissenters from orthodox view was Albert Einstein, who at the Fifth Physical Conference of the Solvay Institute held in Brussels, 1927, presented several arguments designed to demonstrate the reality of unobserved phenomena and hence the incompleteness of quantum mechanics. He found especially disturbing the growing belief among his contemporaries that not only may knowledge of one of a system's physical attributes preclude simultaneous knowledge of others (e.g., momentum and position of a particle), but that they may not simultaneously exist - the principle of complementarity thought to be codified in the Heisenberg uncertainty relations, e.g.
\bea
\Delta x \Delta p & \geq & \hbar \label{1}
\eea
where the uncertainties here measure a quantity's spread about its experimental mean. It was also at the Solvay V meeting that Heisenberg, representing the orthodox view and in open opposition to Einstein, pronounced quantum mechanics a theory full and complete, not subject to further modification. 

\noindent Einstein argued by way of thought experiments, sometimes called gadanken experiments, one of which involved the analysis of a variation on the double-slit experiment. In it the opaque screen in which the double slits are made now moves freely in the vertical plane, sensitive to minute vertical momentum transfer 
\begin{center}
\scalebox{0.3}[0.3]{\includegraphics{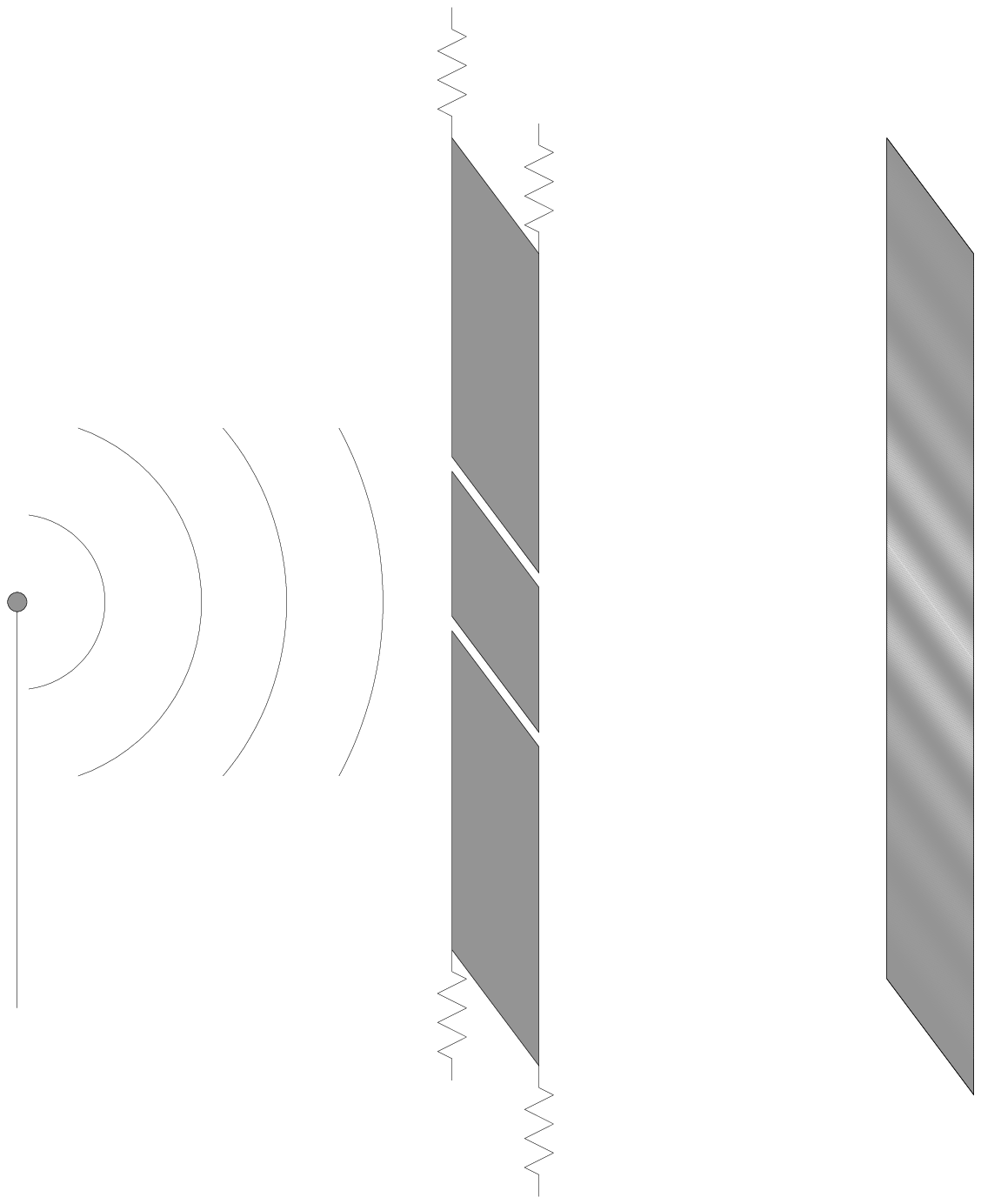}}\\
figure 5
\end{center}
As a horizontally moving electron clears a slit it acquires vertical momentum, which from momentum conservation is balanced by the opposite vertical momentum of the opaque screen. The screen's momentum might then be measured to arbitrary accuracy by an attached meter, thus reducing the momentum spread, $ \Delta p $, indefinitely, independent of the position spread at the observation screen, $ \Delta x $,  in violation of (\ref{1}). 

\noindent Leading the orthodox side at the conference was Niels Bohr of the University of Copenhagen where much of the quantum mechanics conceptual work had been done, whence the name ``Copenhagen interpretation'' often used in reference to the orthodox view. To this argument Bohr cautioned that not only do microscopic objects obey quantum mechanics, but macroscopic objects also. If, e.g., the attached measuring device is capable of resolving upper from lower slit momentum transfer for an electron that eventually lands on a given bright fringe, then classically we must have
\bea
\Delta p_s \leq | p_u - p_l | & \simeq & h \nu | \theta_u - \theta_l | \simeq (h a )/(\lambda d ) \nonumber 
\eea
and for the opaque screen, from (\ref{1}), quantum mechanically 
\bea
\Delta x_s \geq h/\Delta p_s \geq (\lambda d)/a
\eea
which is just the separation between observation screen lines. The opaque screen position spread accordingly affects that measured at the observation screen, $ \Delta x $, here to the extent that distinct observation screen lines will not even appear, and so demonstrating a correlation between complementary spreads, $ \Delta x \, \& \, \Delta p $, dramatically.  The Solvay V participants seem generally to have accepted this explanation. Bohr had similar success with the other various thought experiments presented then by Einstein, likewise those presented three years later at Solvay VI \footnote{ The Einstein-Bohr debate, as it has been called, spanned several years, involving many auxiliary players. An excellent historical account can be found in Max Jammer's  { \sl The Philosophy of Quantum Mechanics} New York: John Wiley \& Sons, Inc. (1974) }. 

\noindent Still sure of his intuition, two years after Solvay VI Einstein in collaboration with Podolsky and Rosen made another attempt to demonstrate the incompleteness of quantum mechanics. The result has been called the EPR paradox  \cite{1}. Key to the approach is the meaning given to the term ``physically real'' for which they propose the sufficient condition
\begin{quotation}
\small 

    \noindent If without in any way disturbing a system we can predict with certainty (i.e., with 
    probability equal to unity) the value of a physical quantity, then there exists an element 
    of physical reality corresponding to the physical quantity.

\end{quotation}
With this definition they then analyze the case of an isolated bound state particle that decays into two identical mass constituents
\begin{center}
\scalebox{0.85}[0.85]{\includegraphics{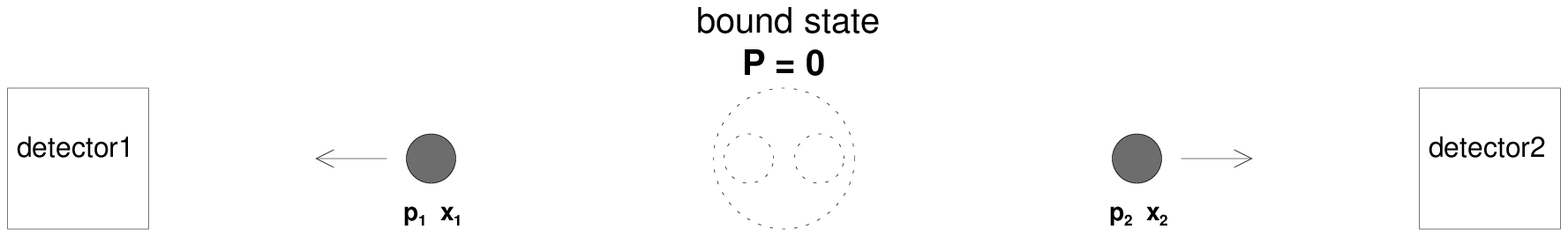}}\\
figure 6
\end{center}

$ \bP = \bp_1 + \bp_2 = 0 $

$ \bX = \bx_1 + \bx_2 = 0 $

\noindent Strict conservation for the isolated system, their argument begins, yields $ \bp_1 $ upon a momentum measurement on particle-2 without in any way disturbing particle-1. To insure this required measurement-independence the momentum measurement is assumed to take place at a distance remote from particle-1, a space-like distance which by special relativity prevents a causal relation between the two events (barring, in the words of Einstein, some ``spooky action-at-a-distance''). Momentum $ \bp_1 $  is thereby guaranteed to be real. But the position of particle-2, the argument continues, could just as well have been measured instead of its momentum, again without in any way disturbing particle-1. The position of particle-1, $ \bx_1 $, is therefore also real. In this way the quantum mechanical description of reality in which the pair ($ \bx_1 , \bp_1 $) is complementary and so do not be simultaneously exist is shown to be logically incomplete.  

\noindent  The resolution of the paradox eventually offered by Bohr within the framework of the Copenhagen view is generally thought to be less convincing than those of his earlier successes. The paradox follows from two main assumptions:
\begin{enumerate}

\item Elements of reality are measurement independent (i.e., measurements only reveal a
    pre-existing reality ), the reality assumption.     
                                            
\item Information cannot be exchanged between space-like separated events, the Locality
    assumption, sometimes called the assumption of Einstein Separability.

\end{enumerate}
The present orthodox view indirectly predicts a violation of assumption 2 for the example above by way of an instantaneous global change in the total two-particle wave function (called a ``collapse'') upon measurement on either of the particles. A test of the assumption would entail an analysis of space-like separated pair-observations, with correlations going to support the orthodox view, and their absence, the EPR view.

\section{Bell inequality}

The EPR experiment as described above would be difficult to carry out in practice, not least because the two measurements would have to be made simultaneously due to the steady spreading (dispersion) of the particles' wave-packets. And from the outset few if any physicists gave serious thought to conducting such an experiment. Then in 1951 David Bohm proposed a more hopeful version of the paradox \cite{2} in which the complementary quantities to be measured are particle spin-projections along distinct axis 
\begin{center}
\scalebox{0.85}[0.85]{\includegraphics{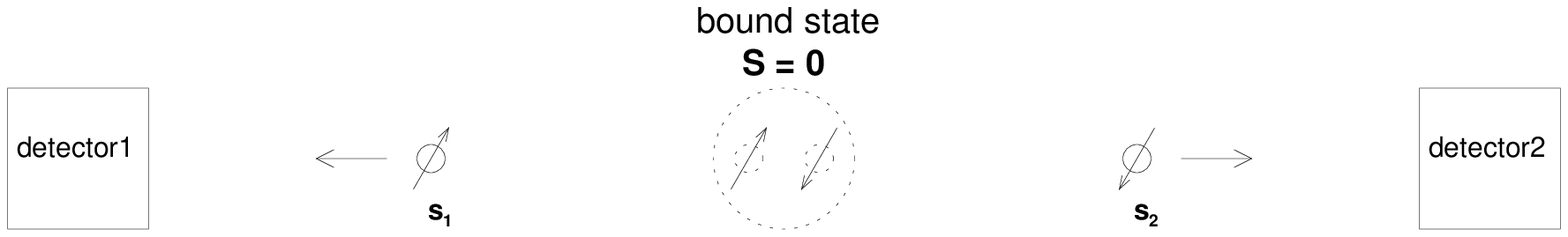}}\\
figure 7
\end{center}
\noindent In this case the pair-wise measurements need not be made simultaneously. The empirical facts that have come to be known are 

\begin{enumerate}

\item Measurement of an individual particle spin along any axis yields one of only two possible outcomes, either spin-up or spin-down.

\item When the two detectors measure along the same axis the outcomes are invariably opposite, one spin-up, the other spin-down (perfect anti-correlation).

\item When the two detectors measure along different axis the outcomes are mixed, sometimes equal, sometimes opposite.

\end{enumerate}

\subsection{Joint-distribution derivations}

Let us consider then a large number of such runs with projections measured randomly along the three directions, $ \hat{\ba}, \hat{\bb}, \& \hat{\bc} $,
\begin{center}
\scalebox{0.4}[0.4]{\includegraphics{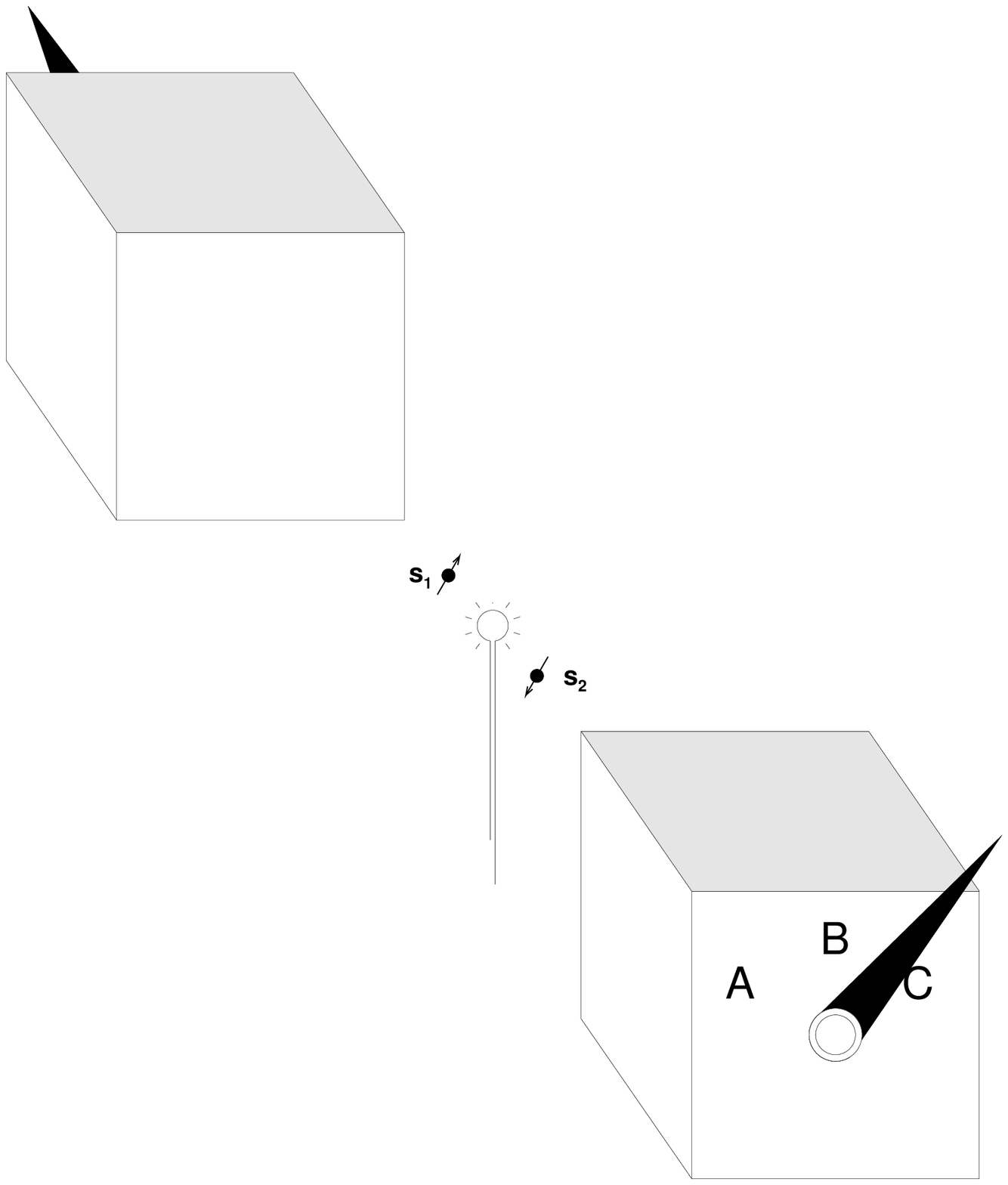}}\\
figure 8
\end{center}
\noindent selected respectively by detector settings A, B, and C, in the end grouping like results. By $ N(a_+ ; b_+ )_{exp} $ we shall designate the number of pair-measurements taken with detector-1 set to direction $ \ba $ and detector-2 to $ \bb $, both yielding spin-up. In the realist view, such as Einstein's, each decay state counted in $ N(a_+ ; b_+ )_{exp} $  had before measurement the property that particle-1 would measure spin-up along direction $  \ba $ and particle 2 along direction $ \bb $. Accordingly, $ N(a_+ ; b_+ )_{exp} $ is also the number of pre-observation states that would later, upon measurement, be observed to have spin-up in directions $ \ba $ and $ \bb $  respectively for particles 1 and 2. To distinguish the pre-observation frequencies we omit the subscript, though the two numbers are of course equal. 
\bea
N(a_+ ; b_+ ) &=&  N(a_+ ; b_+ )_{exp} \label{2}
\eea
For the pre-observation states, however, we may also write

\bea
N(a_+ ; b_+ ) &=&  N(a_+ ; b_+, c_+ ) + N(a_+ ; b_+, c_- )  \nonumber \\ &=&
  N(a_+ , c_+; b_+ ) + N(a_+ , c_-; b_+ )    \label{3}
\eea
from completeness, where the second index on either side of the argument semicolon gives the experimental outcome in case the projection were measured in a different direction ( above, direction $ \bc $ ) from that given by the first index. In this notation the semicolon separates particle-1 from particle-2 projections, commas separating the projections within the set of particle-1 or particle-2 projections. The first particle-1 and particle-2 indices give the physical detector setting for that measurement, the second, where there is one, a hypothetical setting. The expansion follows from the reality assumption-1 above.

\noindent For the measurements considered in fig (8) we have then,
\bea
N(a_+ ; b_+ ) &\geq &  N(a_+ ; b_+, c_+ ) = N(a_+ ; c_+, b_+ )  \label{4}
\eea
and
\bea
N(b_+ ; c_+ ) &\geq &  N(b_+ ; c_+, a_- ) = N(a_+ ; c_+, b_- )  \nonumber
\eea
so that
\bea
N(a_+ ; b_+ ) + N(b_+ ; c_+ ) &\geq &  N(a_+ ; c_+, b_+ ) + N(a_+ ; c_+, b_- )  \nonumber
\eea
or
\bea
N(a_+ ; b_+ ) + N(b_+ ; c_+ ) &\geq &  N(a_+ ; c_+) \label{5}
\eea
This is one form of a Bell Inequality (BI) \cite{3}. 

\subsection{Elementary set-theory derivations}

For a different derivation we first group the particles to be observed at detector-1, particle-1s, into sets according to their positive spin projections. Let set A consist of all such particle-1s with projection spin-up along direction $ \ba $. Likewise sets B and C for directions $ \bb $ and $ \bc $ , respectively. From elementary set algebra then we have
\bea
A \cap \bar{B} \cup B \cap \bar{C} & \supseteq & A \cap \bar{C} \label{6}
\eea
giving

\bea
\eta ( A \cap \bar{B} ) + \eta (  B \cap \bar{C} ) & \geq & \eta (A \cap \bar{C} ) \label{7}
\eea
where $ \eta $ is a cardinality operator that counts the number of set elements. In terms of the number of particle-1s with given spin projections we have
\bea
\eta ( A \cap \bar{B} ) &=&  N(a_+, not-b_+)   \nonumber
\eea
where now both N arguments (separated by a comma) are particle-1 projections. This yields from (\ref{7})
 \bea
N(a_+, not-b_+)   + N(b_+, not-c_+)   & \geq & N(a_+, not-c_+)   
\eea
from which observed spin conservation obtains
 \bea
N(a_+; b_+)   + N(b_+; c_+)   & \geq & N(a_+; c_+)   \label{8}
\eea
where by convention the semicolon again separates particle-1 from particle-2 projections. A Bell inequality \cite{4}.

\subsection{Expectation-value derivations}

In quantum mechanics $ N(a_+; b_+) $ is proportional to the joint probability of measuring particle-1 spin-up along direction $ \ba $ and particle 2 spin-up along direction $ \bb $ 
\bea
n(a_+; b_+) &=& \lan 0,0 | \frac{1}{2} ( 1 + \bsigma_1 \cdot \ba  )  \frac{1}{2} (1 + \bsigma_2 \cdot \bb  )  | 0,0  \ran    \label{8.5} \\ &=&
\frac{1}{2} \sin^2(\frac{1}{2} \theta_{ab} ) \nonumber
\eea
Bell's inequality often appears in the literature as a relation between expectation values. The singlet-state product spin projection expectation for particles 1 and 2 along directions $ \ba $ and $ \bb $, respectively, is given by
\bea
P(\ba; \bb) &=& \lan 0,0 | \bsigma_1 \cdot \ba  \bsigma_2 \cdot \bb   | 0,0  \ran   = 4 n (a_+; b_+ ) - 1 \nonumber
\eea
yielding from (\ref{8}) above
\bea
P(\ba; \bb) + P(\bb; \bc) & \geq & P(\ba; \bc) -1 \label{9}
\eea
\cite{5, 38}. It happens that quantum mechanics predicts a violation of the Bell Inequality. From (\ref{8}) 
\bea
\sin^2(\frac{1}{2} \theta_{ab} ) + \sin^2(\frac{1}{2} \theta_{bc} ) & \geq & \sin^2(\frac{1}{2} \theta_{ac} ) \nonumber
\eea
which is violated e.g. for relative detector settings $ \theta_{ab} = \theta_{bc} = \pi/8, \theta_{ac} = \pi/4 $ . These give
\bea
2 \sin^2( \pi/8 ) & \geq & \sin^2(\pi/4 ) \nonumber \\ 
.29 & \geq & .5 \nonumber
\eea
The predictions of quantum mechanics and its realist interpretation ( leading from assumptions 1 and 2 above ) appear to be at variance. The experimental data to date tends to agree with quantum mechanics.

\section{The inequality reconsidered}

We now take a closer look at the derivations and consider a couple of others. Recall that the inequality relates classes of EPR experimental data where detectors 1 and 2 of figure(8) are set to all combinations of directions $ \ba, \bb,$ and $ \bc $ for an indefinitely large number of measurements. For each individual detector setting the measurement result is either spin-up or spin down. We designate by $ \Omega $ the set of all pre-observation particle-pair states to be measured, and by A, B, and C, as before, the sets of states in $ \Omega $  in which particle-1 has positive spin projection along directions $ \ba, \bb,$ and $ \bc $, respectively. We then have again relation (\ref{6}) 
\bea
\eta ( A \cap \bar{B} ) + \eta (  B \cap \bar{C} ) & \geq & \eta (A \cap \bar{C} ) \nonumber
\eea
which we illustrate here by Venn diagram
\begin{center}
\scalebox{0.5}[0.5]{\includegraphics{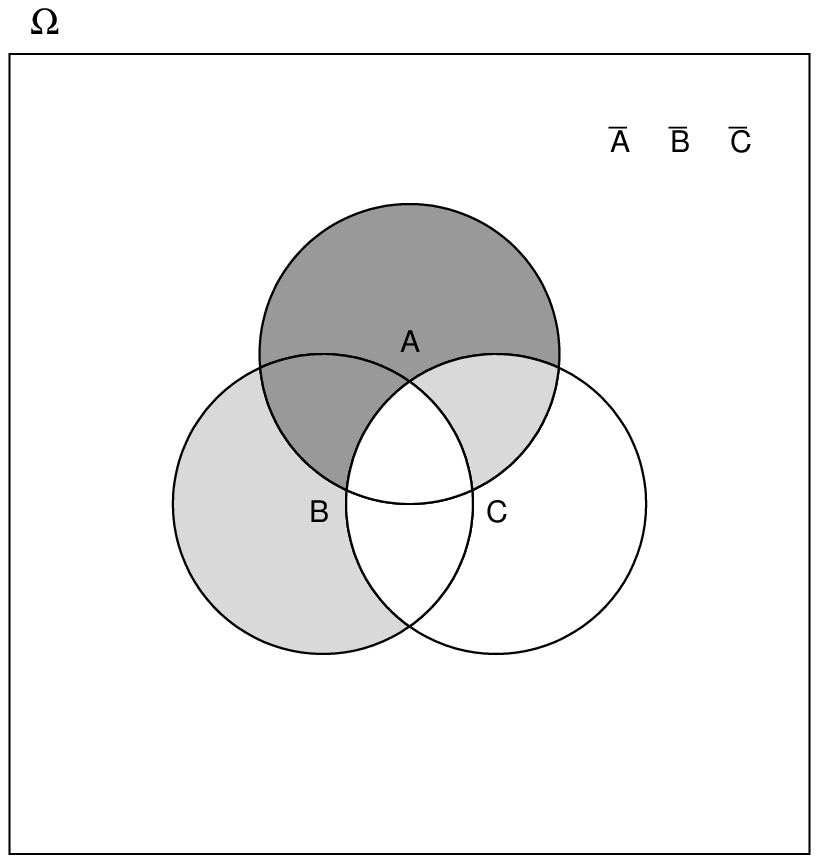}}\\
figure 9
\end{center}
or in terms of Venn sections
\begin{center}
\scalebox{1.3}[1.3]{\includegraphics{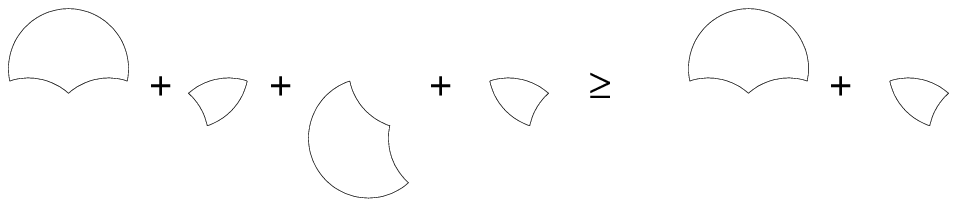}}\\
figure 10
\end{center}
By $ \tilde{N}(a_+; b_+) $ we designate the number of states in $ \Omega $ that have positive spin projections along $ \ba $ and $ \bb $ for particles 1 and 2 (of a pair), respectively. Then
\bea
\tilde{N}(a_+; b_+)   &=& \eta ( A \cap \bar{B} )  \nonumber
\eea
yielding from cardinality relation (\ref{7})
\bea
\tilde{N}(a_+; b_+)   + \tilde{N}(b_+; c_+)   & \geq & \tilde{N}(a_+; c_+)   \label{10}
\eea
an inequality similar to Bell's (\ref{8}). 

\noindent To investigate the relation first recall that an expression $ N(\alpha_+;\beta_+) $ appearing in (\ref{8}), like its experimental counterpart $ N(\alpha_+;\beta_+)_{exp} $,  represents the number of states in $ \Omega $  that have positive spin projections along $ \balpha $ and $ \bbeta $ for particles 1 and 2 (of a pair) that {\em will be} observed at detectors 1 and 2 along directions $ \balpha $ and $ \bbeta $ , respectively. The set of pairs counted in $ N(\alpha;\beta) $ is therefore only a subset of the set of pairs in the corresponding $ \tilde{N}(\alpha;\beta) $; some of those in $ \tilde{N}(\alpha;\beta) $ may be counted in $ N(\alpha^\prime ;\beta^\prime) $.  Then let $ \Omega_{\alpha \beta } \subseteq \Omega $  designate the set of states in $ \Omega $ and $ A_{\alpha \beta } \subseteq A $ the set of states in A upon which detectors 1 and 2 eventually measure spins along directions $ \balpha $ and $ \bbeta $, respectively, so that N bears the same relation to $ \Omega_{\alpha \beta } $ and $ A_{\alpha \beta } $ as $ \tilde{N}$ to $ \Omega $ and A. In terms of the three quantities with observed counterparts, $ N(a_+; b_+) , N(b_+; c_+), $ and  $ N(a_+; b_+) $ , we therefore have in the double entry notation for $\tilde{N}$.
\bea
\tilde{N}(a_+ ; b_+ ) &=&  N(a_+ ; b_+ ) + N(a_+ ; c_+, b_+ )  \nonumber \\
\tilde{N}(b_+ ; c_+ ) &=&  N(b_+ ; c_+ ) + N(a_+ , b_+; c_+ )  \nonumber \\
\tilde{N}(a_+ ; c_+ ) &=&  N(a_+ ; c_+ ) + N(a_+ ; b_+, c_+ )  + N(b_+ , a_+; c_+ )  \nonumber
\eea
which upon substitution into inequality (\ref{10}) yields
\bea
\lefteqn{ N(a_+; b_+) + N(b_+; c_+) \geq N(a_+; c_+) } \label{11} \\
 && + [N(a_+ ; b_+, c_+ )  - N(a_+ ; c_+, b_+ )   ] + [N(b_+ , a_+; c_+ )  - N(a_+ , b_+; c_+ )   ]  \nonumber
\eea
reducing to BI (\ref{8}) on the condition that
\bea
N(a_+ ; b_+, c_+ )  + N(b_+ , a_+; c_+ )  &=& N(a_+ ; c_+, b_+ )  + N(a_+ , b_+; c_+ )   \label{12}
\eea
or equivalently, from spin conservation, on the condition that
\bea
N(a_+ ; b_+, c_+ )  + N(b_+ ; c_+, a_- )  &=& N(a_+ ; c_+, b_+ )  + N(a_+ ; c_+, b_- )  = N(a_+ ; c_+ )   \nonumber  
\eea
the hidden assumption, equation (\ref{4}), which led to Bell inequality (\ref{5}). Justification of this assumption is main the subject of the remainder of this article.

Let us consider the assumption first by way of various Venn diagrams, keeping in mind the singlet state condition, $ \tilde{N}(\alpha; \alpha, \beta) = 0 $
\begin{center}
\scalebox{0.8}[0.8]{\includegraphics{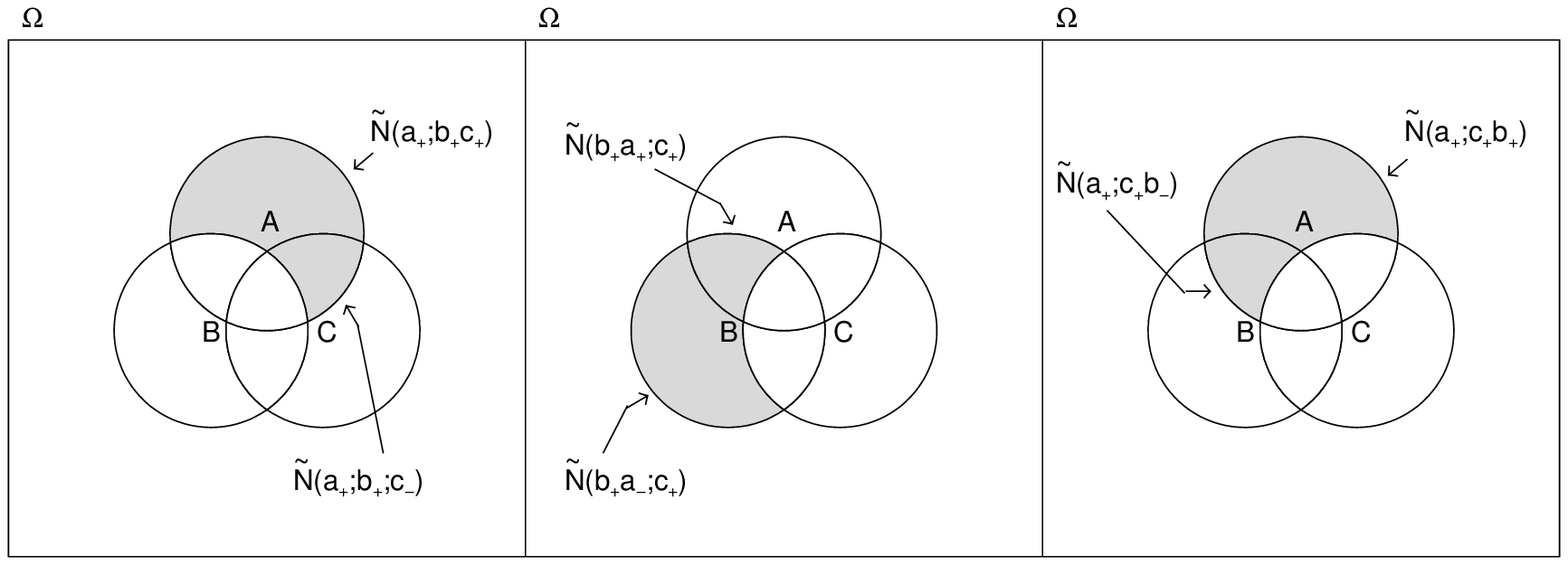}}\\
figure 11
\end{center}
Observe that A B and C overlap, and that
\bea
\eta( A \cap \bar{B} \cap \bar{C}  ) &=& \tilde{N}(a_+ ; b_+, c_+ )  = \tilde{N}(a_+ ; c_+, b_+ )  \label{13}
\eea
and
\bea
\eta( A \cap B \cap \bar{C}  ) &=& \tilde{N}(b_+ , a_+; c_+ )  = \tilde{N}(a_+ , b_+; c_+ ) \nonumber
\eea
By contrast, for the three disjoint subscripted sets $ \Omega_{ \alpha \beta } $ relevant to quantities appearing in (\ref{8})
\begin{center}
\scalebox{0.8}[0.8]{\includegraphics{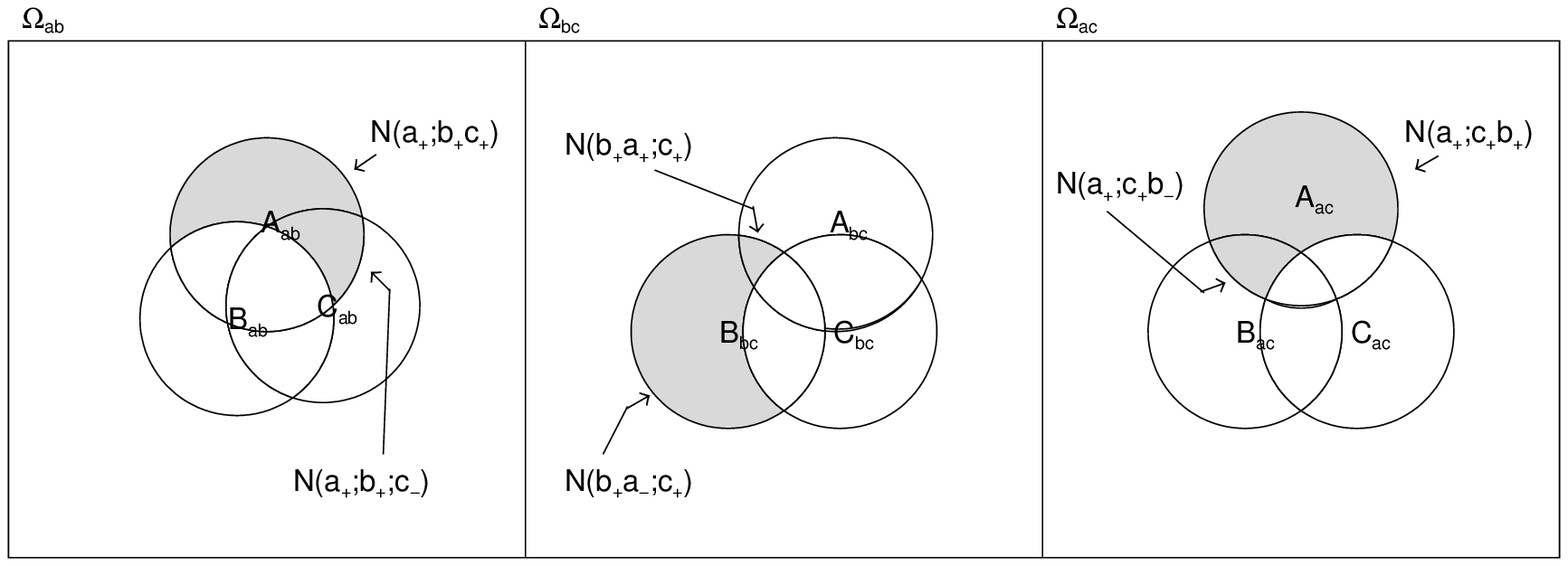}}\\
figure 12
\end{center}
as the distributions are generally distinct. Then, generally
\bea
\lefteqn{ \eta( A_{ab} \cap B_{ab}^\prime \cap C_{ab}^\prime  ) = N(a_+ ; b_+, c_+ ) } \label{14} \\  && \qquad \qquad \qquad \qquad \qquad \neq  N(a_+ ; c_+, b_+ )  = \eta( A_{ac} \cap C_{ac}^\prime \cap B_{ac}^\prime ) \nonumber
\eea
and
\bea
\eta( B_{bc} \cap A_{bc} \cap C_{bc}^\prime  ) = N(b_+ , a_+; c_+ )  & \neq &  N(a_+ , b_+; c_+ )  = \eta( A_{ac} \cap B_{ac} \cap C_{ac}^\prime  ) \nonumber
\eea
This observation contradicts (\ref{4}) and (\ref{12}) upon which derivations (\ref{5}) and (\ref{8}) depend. The skewed spin-projection distributions pictured in the Venn diagrams (12) we illustrate alternatively by superimposing on the $\Omega$ sample space of diagram (11)
\begin{center}
\scalebox{0.5}[0.5]{\includegraphics{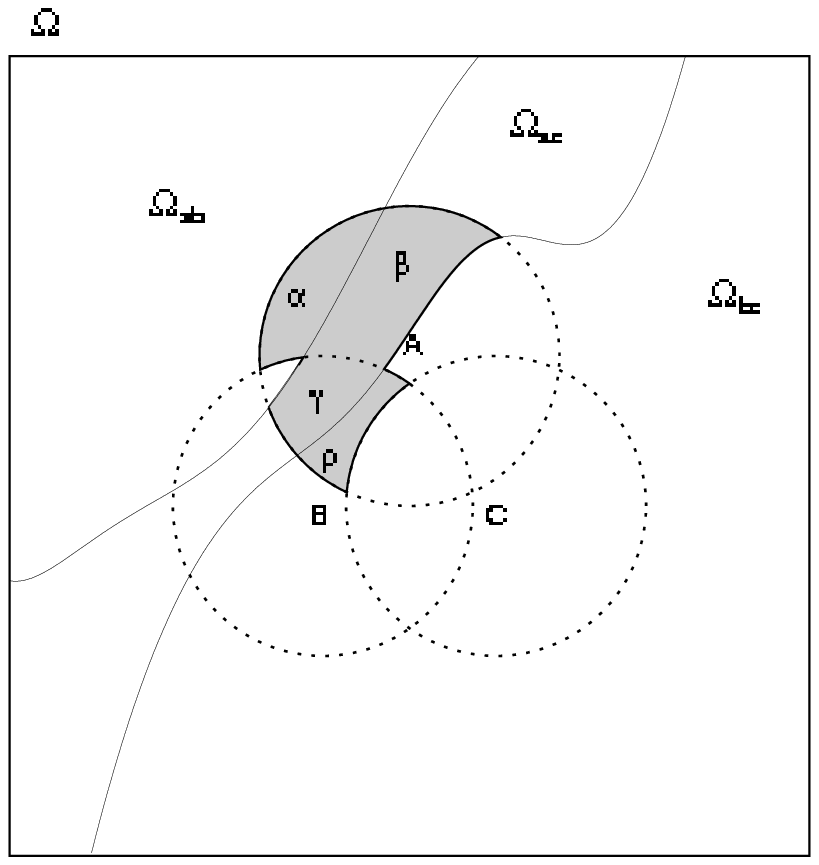}}\\
figure 13
\end{center}
where
\bea
\alpha & \equiv & A_{ab} \cap B_{ab}^\prime \cap C_{ab}^\prime  \nonumber \\
\beta & \equiv & A_{ac} \cap C_{ac}^\prime \cap B_{ac}^\prime  \nonumber \\
\gamma & \equiv & A_{ac} \cap B_{ac} \cap C_{ac}^\prime  \nonumber \\
\rho & \equiv & B_{bc} \cap A_{bc} \cap C_{bc}^\prime  \nonumber
\eea
Contrasting relations (\ref{13}) and (\ref{14}) may also be understood in terms of the N and $ \tilde{N} $ number notation itself. From construction, the first particle indices of N alone refer to actual observations and may therefore not be transposed without further assumptions.

\subsection{ Random-selection derivations }

One such assumption, or condition, is that detector directions be chosen from $ \ba, \bb, $ and $ \bc $ randomly. This approach has the added advantage of avoiding the suspect incompatible joint probabilities \cite{6} $ N(\alpha; \beta \gamma) $ and $ N(\alpha , not-\beta) $ used above. Under this assumption sample space $ \Omega $ of fig (13) approaches homogeneity in $ \Omega_{ab}, \Omega_{bc}, $ and $ \Omega_{ac}  $ which may be pictured
\begin{center}
\scalebox{0.5}[0.5]{\includegraphics{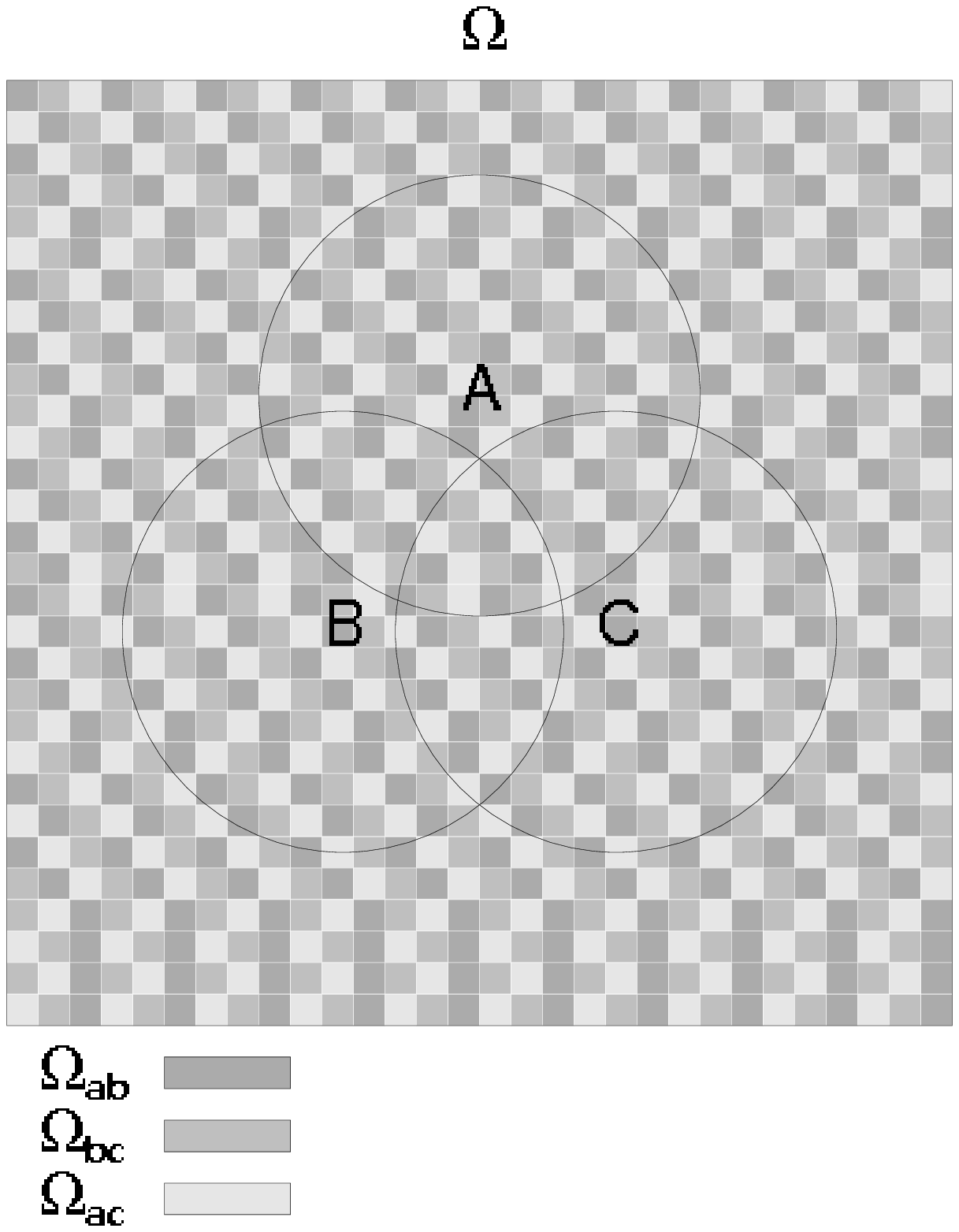}}\\
figure 14
\end{center}
from which, visually
\bea
N(a_+ ; b_+, c_+ )  + N(b_+ ; c_+, a_- )  &\simeq & N(a_+ ; c_+, b_+ )  + N(a_+ ; c_+, b_- )  \nonumber , 
\eea
for the case of interest where $ \eta(\Omega_{ab}) \simeq \eta( \Omega_{bc}) \simeq \eta( \Omega_{ac})  $.

\noindent This randomness is attained on the condition that the observed relative frequencies $ n(\alpha; \beta ) \propto N(\alpha; \beta )  $ remain invariant under variation in place-selection, the order or sequence in which pair projections along the various directions are measured. When this is achieved we have
\bea
n(\alpha_+; \beta_+ ) &=& \tilde{n}(\alpha_+; \beta_+ )  \label{15}
\eea
where $ \tilde{n}(\alpha; \beta ) \propto \tilde{N}(\alpha; \beta )  $, so that
\bea
N(\alpha_+; \beta_+ ) &\propto & \tilde{N}(\alpha_+; \beta_+ )  \nonumber
\eea
which from (\ref{10})  yields the Bell Inequality \cite{6}.
 \bea
N(a_+; b_+)   + N(b_+; c_+)   & \geq & N(a_+; c_+)   \label{16}
\eea
From (\ref{15}) also follows equations (\ref{4}) and (\ref{12}), and hence Bell inequality (\ref{5}) and (\ref{8}).

\subsection{Stochastic hidden-variable derivations}

The biased spin-projection distribution pictured in fig(13) is sometimes described by classical local hidden variables \cite{18, 19, 5, 35, 36, 38} collectively denoted $ \lambda $
\begin{center}
\scalebox{0.5}[0.5]{\includegraphics{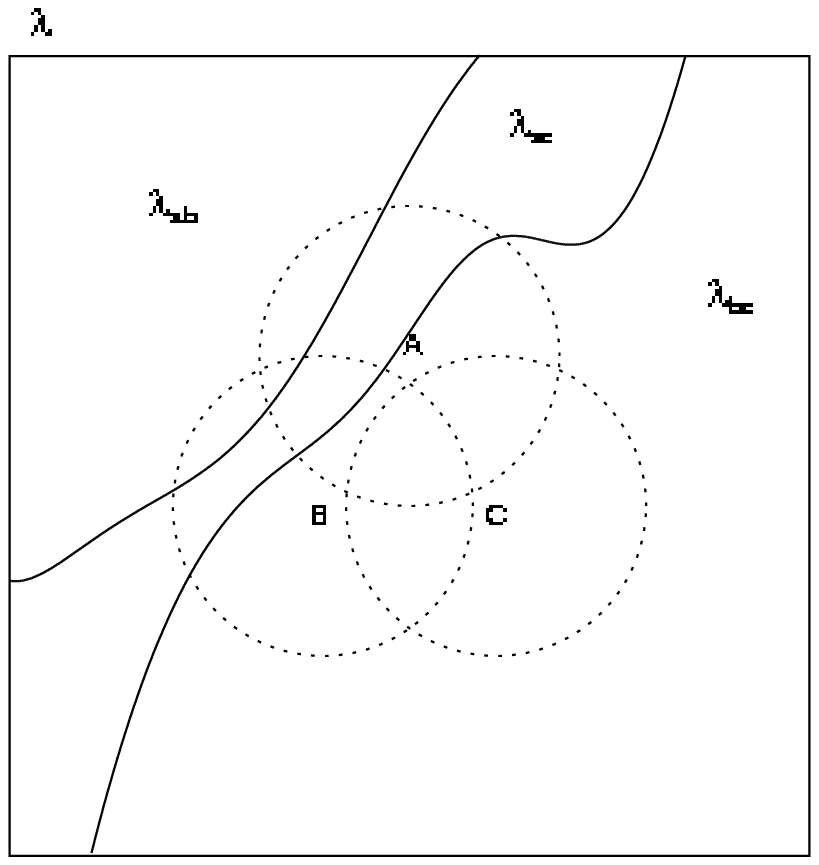}}\\
figure 15
\end{center}
where the states measured along directions $\balpha $ and $ \bbeta $ for particles 1 and 2, respectively( i.e., those in $\Omega_{\alpha \beta } ) $ are characterized by hidden variables $ \{ \lambda_{\alpha \beta } \} $ .
with accompanying classical probability densities
\bea
\rho(\lambda^\prime) &=& \rho_{\alpha \beta}  \qquad \qquad for \qquad \lambda^\prime \in \{ \lambda_{\alpha \beta} \}. \nonumber
\eea
Predictions for measured frequencies then take the form (cf. (\ref{8.5}))
\bea
n_{exp}(\alpha_+; \beta_+ ) &=& n(\alpha_+; \beta_+ )  \nonumber \\ &=&
N_{\Omega_{\alpha \beta }}\int_{\Omega_{\alpha \beta }} d\lambda \rho (\lambda) \lan 0,0; \lambda | \frac{1}{2} ( 1 + \bsigma_1 \cdot \balpha  )  \frac{1}{2} (1 + \bsigma_2 \cdot \bbeta  )  | 0,0; \lambda  \ran    \nonumber
\eea
where the integration is carried out over sub-ensemble $ \Omega_{\alpha \beta } $, $ N_{\Omega_{\alpha \beta }} $ functioning as a normalization constant. On the other hand
\bea
\tilde{n}(\alpha_+; \beta_+ ) &=&
N_{\Omega}\int_{\Omega} d\lambda \rho (\lambda) \lan 0,0; \lambda | \frac{1}{2} ( 1 + \bsigma_1 \cdot \balpha  )  \frac{1}{2} (1 + \bsigma_2 \cdot \bbeta  )  | 0,0; \lambda  \ran  . \nonumber
\eea
Similarly, we have for the expectation values (cf. (\ref{9})),
\bea
P(\balpha; \bbeta ) &=&
N_{\Omega_{\alpha \beta }}\int_{\Omega_{\alpha \beta }} d\lambda \rho (\lambda) \lan 0,0; \lambda | \bsigma_1 \cdot \balpha  \bsigma_2 \cdot \bbeta    | 0,0; \lambda  \ran . \nonumber
\eea
From the algebraic identity
\bea
-1 & \leq &  a b + b c - a c + b^2- 2 b  \leq 0 \nonumber
\eea
one might say 
\bea
\lefteqn{ - {\bf 1 } \leq \bsigma_1 \cdot \ba \bsigma_2 \cdot \bb + \bsigma_1 \cdot \bb \bsigma_2 \cdot \bc - \bsigma_1 \cdot \ba \bsigma_2 \cdot \bc } \nonumber \\ && \qquad \qquad \qquad \qquad + \bsigma_1 \cdot \bb \bsigma_2 \cdot \bb - \bsigma_1 \cdot \bb - \bsigma_2 \cdot \bb \leq {\bf 0 } \nonumber
\eea
with regard to its expectation value. Operating then from the right with ket $ | 0,0; \lambda \ran $ and from the left with  $ N_{\Omega} d\lambda \rho (\lambda) \lan 0,0; \lambda | $   yields upon integration over $ \Omega $  \cite{19, 34, 35, 36, 37, 41}
\bea
\tilde{P}(\ba; \bb) + \tilde{P}(\bb; \bc) & \geq & \tilde{P}(\ba; \bc) -1 \label{16.5}
\eea
with
\bea
\tilde{P}(\balpha; \bbeta ) &=&
N_{\Omega}\int_{\Omega} d\lambda \rho (\lambda) \lan 0,0; \lambda |  \bsigma_1 \cdot \balpha \, \bsigma_2 \cdot \bbeta   | 0,0; \lambda  \ran  .\nonumber
\eea
which on the stochastic condition by which, again, observed relative frequencies remain invariant under variation in place-selection, yields
\bea
P(\balpha; \bbeta ) &=& \tilde{P}(\balpha; \bbeta ) \label{16.6}
\eea
and Bell Inequality (\ref{9}). 

\noindent The stochastic condition is equivalent to the condition - in the language of the quantum mechanics canonical formalism - that the experimental state-preparation density matrix remain independent of detector settings, the place-selection sequence. I.e., that
\bea
\int d\lambda \rho_{\alpha \beta } | 0,0; \lambda \ran \lan 0, 0 ;\lambda | \simeq \int d\lambda \rho_{\alpha^\prime \beta^\prime } | 0,0; \lambda \ran \lan 0, 0 ;\lambda |  \nonumber
\eea

\section{The Semifactual}
                      
In the usual statistics problem it is safe to assume that successive place-selections are taken over identical or sufficiently similar sequences of sample events, that the sample sequence and associated selection sequence remain mutually independent. This however is not the usual case, and with regard to possibly unknown or ``hidden'' factors the experimentalist is obliged to take special precaution to prevent the transfer of information from selection to sample. The recommended precaution in every case of which I am aware has been that there be established space-like separations between a detector's measurement on a particle and the corresponding opposite particle of the pair whose spin-projection is presently measured at the opposite detector. The opposite particle will then, presumably, have no way of knowing how to adjust its spin direction so as to produce with the other measurements the observed correlations - no way, that is, barring superluminal information transfer. This condition may also be applied directly to expressions $ N(\alpha_+; \beta_+, \gamma_+ ) $ \cite{17, 42} to yield  (\ref{4}) and (\ref{12}). On the other hand, we illustrate the general relation, i.e. a sample dependence on selection, for the $ i^{th} $ selection sequence with the sample sequence given in terms of its hidden variable
\begin{center}
\scalebox{0.8}[0.8]{\includegraphics{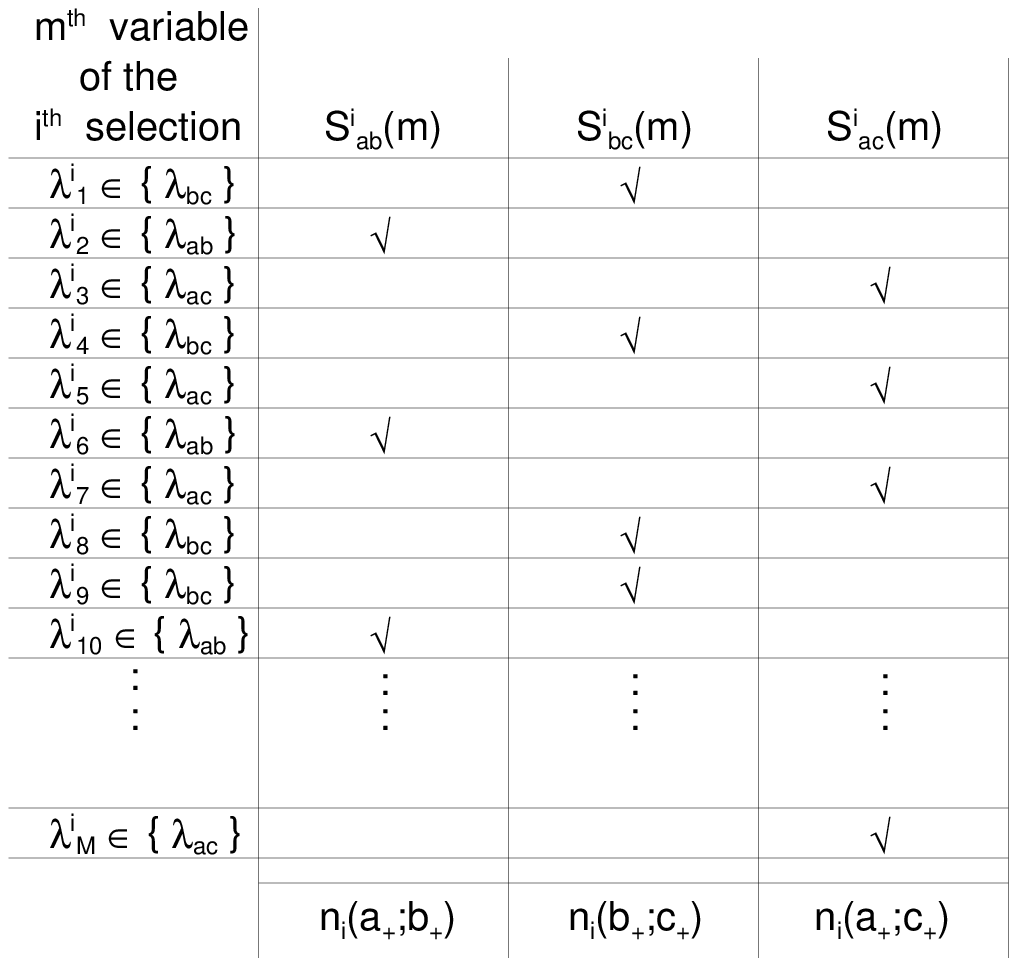}}\\
figure 16
\end{center}
where $ S^i(m) = \{ S^i_{ab}(m), S^i_{bc}(m), S^i_{ac}(m)  \} $ is the $ i^{th} $ selection for the $ m^{th} $ particle (whose hidden variable is $ \lambda_m^i $) with $ S^i_{\alpha \beta}(m) $ ``selected'' and ``not-selected'' indicated  respectively in its column by ``$ \surd $'' and ``  ``. 

\noindent From schematic (16) we may understand the justification of (\ref{4}), (\ref{12}) \& (\ref{15}), leading to Bell inequalities (\ref{5}), (\ref{8}) \& (\ref{16}), in terms of the conditional

\begin{quotation}
\noindent  given $ S^i $ and $ S^j $ \\ 
if  $ n_i(\alpha; \beta)  \simeq  n_j(\alpha; \beta) $ \, and \,  
$ \rho^i_{\alpha \beta} \simeq \rho^j_{\alpha \beta} \,\, \forall \,\, i \& j $ \, (randomness) \\ 
then $  n_i(\alpha; \beta)  \simeq  \tilde{n}_i(\alpha; \beta) \to \tilde{n}(\alpha; \beta) $ \, in the large $ \eta(\Omega) $ limit. 
\end{quotation}
In terms of error analysis, the first part of the antecedent, $ n_i(\alpha; \beta)  \simeq  n_j(\alpha; \beta) $, secures against statistical bias, and the second, $ \rho^i_{\alpha \beta} \simeq \rho^j_{\alpha \beta} $, against systematic bias. With the first part now long established\cite{7}we here consider the second part. And while its violation would involve a remarkable conspiracy on the part of nature, the question cannot be answered empirically since the spin distributions are joint over incompatible observables. It is therefore the absence or impossibility of a selection-source sequence correlation that we take for the affirmative answer, $ \rho^i_{\alpha \beta} \simeq \rho^j_{\alpha \beta}$ .
 
In the usual idealized treatment, briefly described above, the experimentalist secures against the possibility of correlation by delaying the actual direction setting on a given detector till {\em the very last instant } before measurement, sufficiently so that the bit of  selection information, $ S^j(m^\prime) $, the detector direction setting,  cannot possibly reach the relevant source element , the opposite particle of the pair with hidden variable $ \lambda^j_{m^\prime} $, before its spin-projection is measured. As already mentioned, this requires the establishment of space-like separations between detector settings and opposite particles. On this condition it might then be said that had one chosen to measure the spin projection along $\bgamma $ instead of $\bbeta $ for a given place selection, the frequency $ n(\alpha_+; \beta_+, \gamma_+ ) $ would have remained essentially unchanged, the condition illustrated in fig (14). In words,

\begin{quotation}
\noindent  if, detector-2 had been set to direction $ \bgamma $ instead of $ \bbeta $ \hfill (i)\\  
then the hidden variable at particle-1, $  \lambda_{\alpha \beta } $, \\ would not have changed (to $ \lambda_{\alpha \gamma }  $ )
\end{quotation}

This together with the observations $ n_i(\alpha; \beta)  \simeq  n_j(\alpha; \beta) $  then yields
\bea
n(\alpha_+; \beta_+, \gamma_+ ) &=& n(\alpha_+; \gamma_+,\beta_+ ) \qquad in \, the \, limit
\eea

Claim (i) is a conditional with a false antecedent (detector-2 is set to direction $ \bgamma $), a counterfactual, with the added property that its consequent ( the hidden variable is $  \lambda_{\alpha \beta } $ ) is true, making the claim a semifactual. The attendant reasoning is therefore called semifactual.

Recall, this is the same reasoning key to the EPR argument, ``Had the position of particle-2 been measured instead of its momentum, the momentum of particle-1 would not have changed''. It follows from an assumption of causal determinism, such as described e.g. by the conditional, if A then C ( the occurrence of A implying the occurrence of C ), with the verification criteria implicit: When condition A is observed, condition C shall always follow or accompany. The semifactual antecedent likewise refers to the {\em future } rather than the past \cite{40}. For example, given the background
\begin{quotation}
 \noindent The day before the drawing Jason choose lottery  \# 11436. \\
                                The winning \# was 94325
\end{quotation}
the semifactual
\begin{quotation}
\noindent  Had Jason's number been 94325 instead of 11436 \\
                                the winning lottery number would still have been 94325
\end{quotation}
may be restated: If for a future drawing Jason chooses number 94325 given that there are sufficiently similar world conditions as those before his last choice, he will win the jackpot. This is of course a prediction, and one whose failure under semifactual reasoning would mean that the world conditions had not been sufficiently similar to those before, that some important factors, possibly hidden, had been altered. The realist interpretation of quantum mechanics is therefore, via determinism, semifactually definite, where semifactual statements are assigned definite true or false values.

\noindent  Quantum mechanics, on the other hand, is an essentially statistical and non-deterministic theory to the extent that it is not always possible to establish event conditionals of the type, if A then C, e.g. for the events relevant to Young's double-slit experiment. In the very formalism sufficient conditions A such as might imply subsequent events C do not simultaneously appear. Quantum mechanics is, accordingly, semifactually indefinite. To counter the apparent paradox presented by EPR it is explained in the present, revised\cite{16}, orthodox view that at the time of the momentum measurement on a particle-2 there is an effective momentum measurement also on the opposite particle-1 via an instantaneous collapse of the wave function \cite{39}. This forces particle-1 into a state of definite momentum no less than it forces particle-2, and so, by the EPR definition, excluding $ \bp_1 $ as an element of reality. Likewise the position of particle-1, $ \bx_1 $. The view that the only real quantities are measured ones is thus reasserted.

\noindent  We here give a schematic showing roughly the context in which the question of semifactuality enters the motivating discussion regarding the existence of unobserved phenomena
\begin{description}
\item{i)} realist argument:
\begin{description}
\item{a)} locality + observed conservation \\ $ \Rightarrow $ existence of real-unmeasured quantities (by EPR criterion )
\item{b)} locality + observed conservation + counterfactual definiteness  \\ $ \Rightarrow $ existence of simultaneous real-unmeasured quantities \\ $ \Rightarrow $ QM incomplete \& BI ( $ \Rightarrow $ QM incorrect )
\end{description}
\item{ii)} orthodox argument:
\begin{description}
\item{a)} non-existence of unmeasured quantities + observed conservation \\ $ \Rightarrow $ non-locality
\end{description}
\end{description}
Now to consider the truth value of semifactual (i). The antecedent we denote ``A'', and the consequent ``C''. We inquire whether the natural-divergence of the antecedent from the actual course of events, the historical record, together with implicit premises converge, in time, back to the original consequent. The implicit premises here consist of the laws of nature, which we designate L, and world history up to the time of the antecedent, $ W_t $ .To begin, these must form together with the antecedent a self-consistent set $ \{A, L, W_t \}$. A semifactual is then determined to be true when the antecedent is either causally irrelevant or purely-positively relevant to the consequent. For the present case irrelevance is the appropriate relation. Such is the inferential program \cite{8} for evaluating natural-divergence semifactuals.

\noindent To illustrate, consider the world historical event
\begin{quotation} 
\noindent Jason turned on the stove
\end{quotation}
and semifactual
\begin{quotation}
\noindent  Had Jason not turned on the stove  \\                              
The kettle would still have come to a boil
\end{quotation}
Since there does exist a causal relation between turning on the stove and heating the kettle (by the laws of physics) the semifactual is determined to be untrue. However, the semifactal:
\begin{quotation}
                                          \noindent Had Jason not tied his shoe-laces \\
                                          The kettle would still have come to a boil
\end{quotation}
might be deemed true, assuming no unknown or hidden relation between the tying of Jason's shoe-lace and the heating of the kettle.

Careful inspection of (i) however reveals an incompatibility between antecedent and implicit premises of a type described in Kvart's excellent work \cite{8} by way of example:
\begin{quotation} \noindent
                     A piece of butter lies in the middle of a cold, barren expanse, \\
                     with no human beings around, at time {\em t } ( the middle of winter).
\end{quotation}
             The associated counterfactual:
\begin{quotation} \noindent
If this piece of butter had been at a temperature of $ 175^o F$ at time {\em t}, \\
                   it would have started to melt.
 \end{quotation}
\begin{quotation} \noindent
             Given that the expanse is continuously cold at times close to {\em t}, the 
             antecedent-event could not have occurred naturally in accordance with
             the prior history of the world up to time {\em t } and the laws of nature.
\end{quotation}
Similarly, the incompatibility in (i) maintains between antecedent A and the experimental facts of the world leading up to A, $ W_t $; it concerns the timing of detector setting decisions; in practice the decisions are never delayed till the given antecedent time ``t'', {\em immediately preceding the measurement}, but are made at a rather earlier time $ t_o $ as part of the experimental setup. In the famous Orsay experiment \cite{9} e.g. the stochastic place-selectors are pre-set long before experimental runs begin. The world prior to $ t $ therefore, with a selector set to instruct detector-2 to, at the appointed time, measure spins along direction $ \bbeta $, is incompatible with a counter instruction given by antecedent A that the measurements be taken instead along direction $ \bgamma $. And the experimental feasibility of the procedure described by antecedent A is doubtful given the time constraints involved. Semifactual (i), accordingly, lacks self-consistency as regards all EPR experiments to date.
The general inconsistency has been noticed by other authors \cite{43} but never examined formally.

\subsection{Semifactual self-consistency}

An obvious choice for an antecedent $ A_o $ that is both consistent with the facts of experimental procedure and incorporates the effect of A would be that ``the stochastic selector is set differently so as to produce effect A''. It has antecedent time, as required, no later than pre-experimental place-selection-setting time, $ t_o $, and we have with the revised now self-consistent set $ \{ A_o, L, W_{t_o} \}$ the revised semifactual
\begin{quotation}
\noindent  if, the place selection had been altered  \hfill (ii) \\
                          then, the spin-state sequence would not have changed
\end{quotation}
It might now be argued here that we have only replaced one incompatible set of implicit premises with another since the decision to set the place-selector one way or the other must be in an experimentalist's mind before he or she can make the actual setting. Here we simply recognize the experimentalist's freedom to choose indeterministically, i.e., independent of the prior history of the world, $ W_{t_o} $, the usual assumption in EPR analysis \cite{9, 6, 33}. While this freedom is not granted under strict determinism, it is in line with most common sense notions; more, it is an assumption made by EPR in deriving the paradox under investigation \footnote{ On the status of counterfactuals under strict determinism ref\cite{8} argues for their lack of meaning from the necessary incompatibility of A with W and L. W and L however are given in the inferential formalism as sufficient conditions only. And one can easily imagine any number of semi-cyclic mechanical processes for which counterfactual reasoning does make sense, both under strict and common-sense determinism, which in such a case are equivalent. Decisive rather are the causal relations existing between antecedent and events causally relevant to the consequent; the elements {\em necessary} to W. Kvart's classification of counterfactuals under strict determinism as world {\sl de-dicto}, i.e. lacking real-world meaning, here appears spurious, an artifact of over-specified W and L. }.

\noindent  And so to proceed with the evaluation of semifactual (ii) we test for causal relevance understood in the Lorentz sense \cite{9, 33, 35, 18} where mutually irrelevant events are separated by  space-like distances. To this end we assign to statements $ A_o $ and C coordinates of the space-time events they describe. To $ A_o $ we assign ``$ x_o $'' about which the physical setting of the place selection by the experimentalist, e.g., by way of setting a randomizer, is localized. To events C we assign the locus of points \{y\}, about which the sample to be observed is localized during the experimental runs. In this notation causal irrelevance is reduced to the condition
\bea
    (x_o -y)^2 & > &1 \label{19}                                     
\eea

\section{Experimental constraints}
 
From (\ref{19}) let us examine the space-time constraints given by the truth conditions necessary for semifactuals (i) and (ii) in the visually simple case of a photon source at the origin; we illustrate the structure in the lab frame with selectors (coincident with associated detectors) set simultaneously and data taken at a constant rate until time $ t_f $
\begin{center}
\scalebox{0.8}[0.8]{\includegraphics{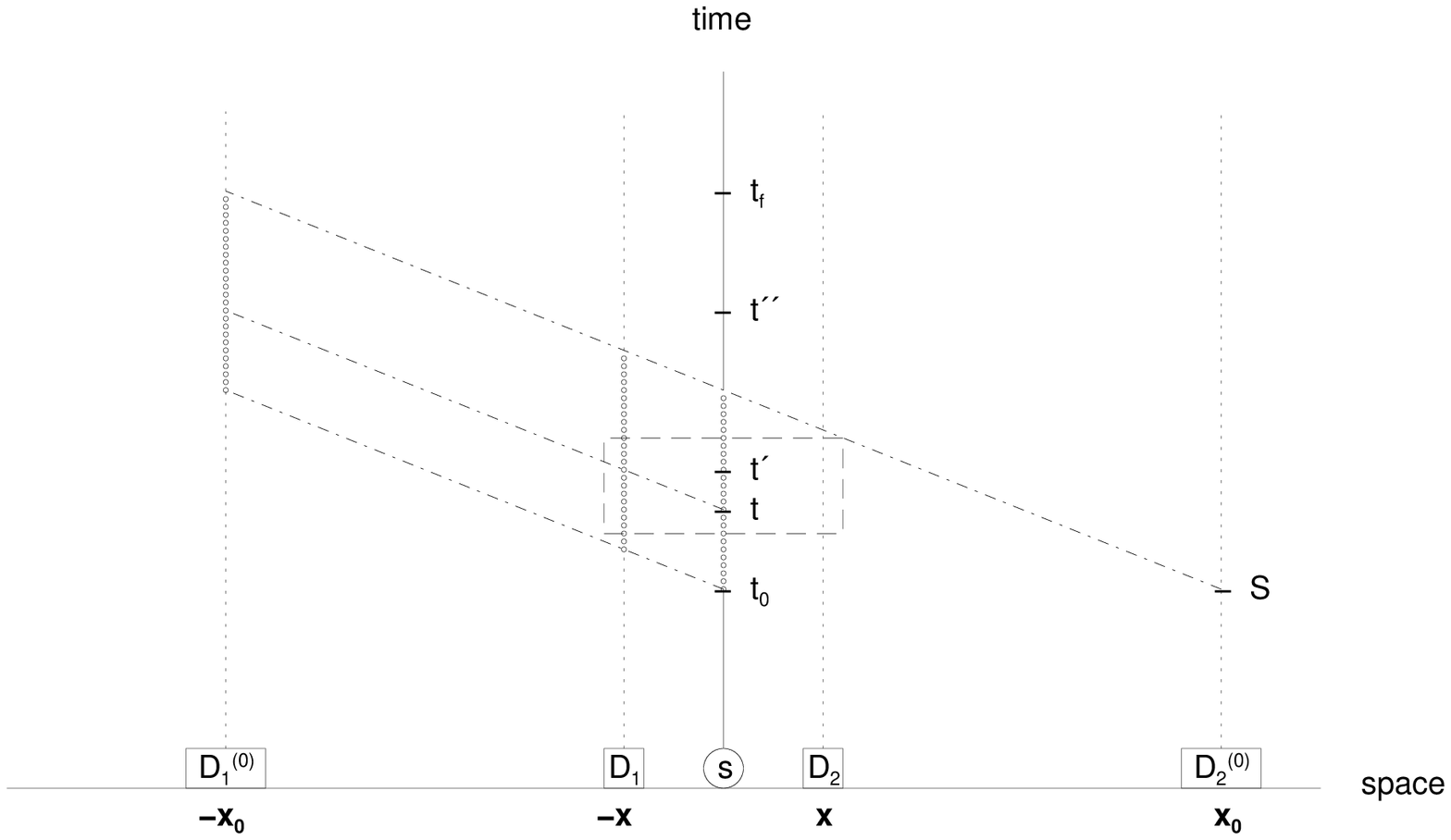}}\\
figure 17
\end{center}
\begin{center}
\scalebox{0.8}[0.8]{\includegraphics{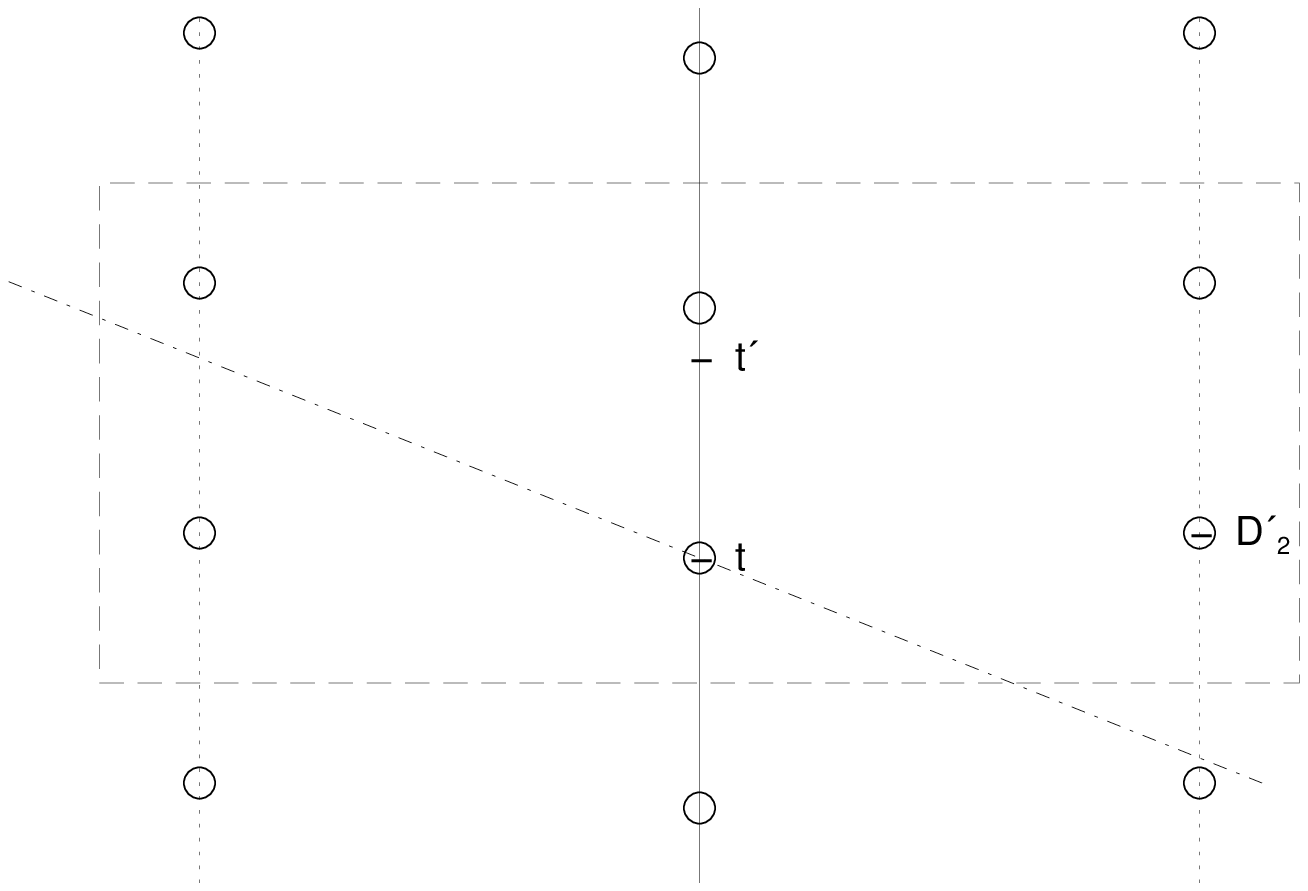}}\\
figure 18
\end{center}
where open dots indicate events - emissions along the source line (sequence events) and detector settings along detector lines (selection events). For the representative photon emitted at time ``t'' the necessary truth condition for (i) is satisfied at detector-1 as the associated detector-2 setting, $ D^\prime_2 $, made { \em at will }by the experimenter lies outside the photon's light-cone. The truth condition for (ii) is likewise satisfied as all pre-determined selections, $ S(t) : t_o \leq t \leq t_f  $, set at time $ t_o $ also lie outside the photon's light-cone.
\noindent Considered in light of the Orsay experimental parameters where the maximum distance between detector and source, $ | \bx | $, is 6.5 meters, the truth condition for semifactual (i) reduces to the requirement that detector-2 settings be made within a maximum of

\bea
\Delta t &=& 4.4 4 10^{-8} seconds    \label{20}
\eea
following each emission to be detected - a time interval that one may compare with the average human reaction time, $ \sim 0.2 $ seconds. 

\noindent Alternatively, for semifactual (ii) the running time of the Orsay experiment, $ t_f - t_o = 12000 seconds $, reduces to a detector separation constraint, $ 2 |\bx| \geq 9.6 \times 10^{12} $ meters, which may be compared with the most widely separated EPR experiment detectors to date, $1.1 \times 10^4 $ meters , or with twice the distance to Neptune, $ 8.7 \times 10^{12} $ meters. From another point of view we might consider the conditions under which real-locality is not violated by the predictions of quantum mechanics. Quantum mechanics predicts for the coplanar angles $ \theta_{ab}= \theta_{bc}= \frac{1}{2} \theta_{ac}= 60^o $ a maximal $ 25\% $ violation of  BI. The violation may therefore be accounted for within the context of real locality when the selection and sequence are mutually localized by the time , $ \tau $, that $ \frac{3}{4}$ the total data has been taken 
\bea
t_o < \tau & \leq & \frac{1}{4} (3 t_f + t_o ) \nonumber
\eea
on the condition 
\bea
L = 2 |\bx - \by| &\leq & \frac{3}{4} (t_f - t_o ) c \nonumber
\eea
where c is the speed of light. For the Orsay experiment with, again, runs lasting 12000s the condition reduces to
\bea
L & \leq & 2.7 \times 10^{12} m \label{21}
\eea
The condition has been met by every EPR experiment to date. Even if the place-selection decision is made and physically entered into the EPR apparatus by the experimentalist within one second of the final measurement, then from the above constraint at least one spatial dimension of the apparatus must be on the order of $ 3 \times 10^8 $ meters. The longest dimension of the Orsay apparatus turns out to be about 12 meters, and the largest of any to date is about 11 kilometers \cite{12}.

\noindent Notice that the above stochastic criteria does not call for "randomness" in the usual sense of effective indeterminacy. For such effectiveness is fundamentally subjective, depending upon subjective discernment. It is required rather that the place-selections and state-sequences remain uncorrelated via causal irrelevance. I.e., the selection-sequence non-correlation necessary for condition (\ref{15}) leading to the BI properly depends upon objective causal irrelevance rather than subjective {\em effective } indeterminacy. To illustrate, consider again semifacual (i), only now imagining as pictured in fig (1) that antecedent A is experimentally feasible, the experimentalist's choice of detector settings being made at will while particles are in mid-flight. In this case the set $ \{A, L, W_t \} $ is perfectly self-consistent and (i) is determined to be true even if detector direction choices are "predictable" with settings e.g. repeating at regular intervals; at any point during such a run the experimentalist is free to break the regularity. And so antecedent A, ``detector-2 is set to direction $ \bgamma $ instead of  $ \bbeta $ ``, would pose no contradiction to the world leading up to the antecedent time, a world in which detector settings had been chosen, $ \bgamma, \bbeta, \bgamma, \bbeta, \bgamma, \bbeta, \bgamma, ... $ or any other way, e.g. ``randomly'' from the spins of a roulette wheel. As illustrated in fig(18) there is no necessary correlation between detector selection $ D^\prime_2 $ and selections vertically below within the photon's backward light cone. The same holds with regard to regularly recurring selections in connection with semifactual (ii). Accordingly, there is no objection to the highly predictable detector setting program implemented by the Orsay experimentalists, and their self-criticism in this regard \cite{10} together with the recent Geneva ``improvements'' \cite{12} appear superfluous \footnote{ In another recent effort an Austrian group \cite{43} employs a "physical" and hence "completely unpredictable" randomizer ( as opposed to a pseudo-random number generator, its workings, however complex, being fundamentally deterministic), proposing to demonstrate indeterminacy via Bell's inequality by first, it would seem, assuming it.}. The EPR detector randomizing problem as conventionally posed is a pseudo problem.

\noindent It is possible to navigate somewhat between the two cases envisaged in semifactuals (i) and (ii) from which the difficult temporal and spatial constraints, (\ref{20}) and (\ref{21}), are derived. One might consider as a starting point the limiting frequency at which indeterminacy can be introduced into the experiment, the inverse human reaction time of  0.2 seconds. E.g., let pre-arranged place-selections be themselves selected from a large varied set {\em at will } by the experimenter at a frequency on the order of the inverse human reaction time. In this case too, as with constraints (\ref{20}) and (\ref{21}), causal irrelevance between selection and sequence is attained. And the associated semifactual has the overall structure of (i) with the substructure of (ii) in the sense that measurements referred to in (i) are replaced by sub place-selections satisfying (ii). The semifactual might read
\begin{quotation}
\noindent if, the sub place-selection had been chosen differently  \hfill (22) \\
                          then, the spin-state sequence would not have changed
\end{quotation}
where sub place-selections are made every 0.2 s. The truth conditions for (22) we illustrate in the space-time diagram
\begin{center}
\scalebox{0.8}[0.8]{\includegraphics{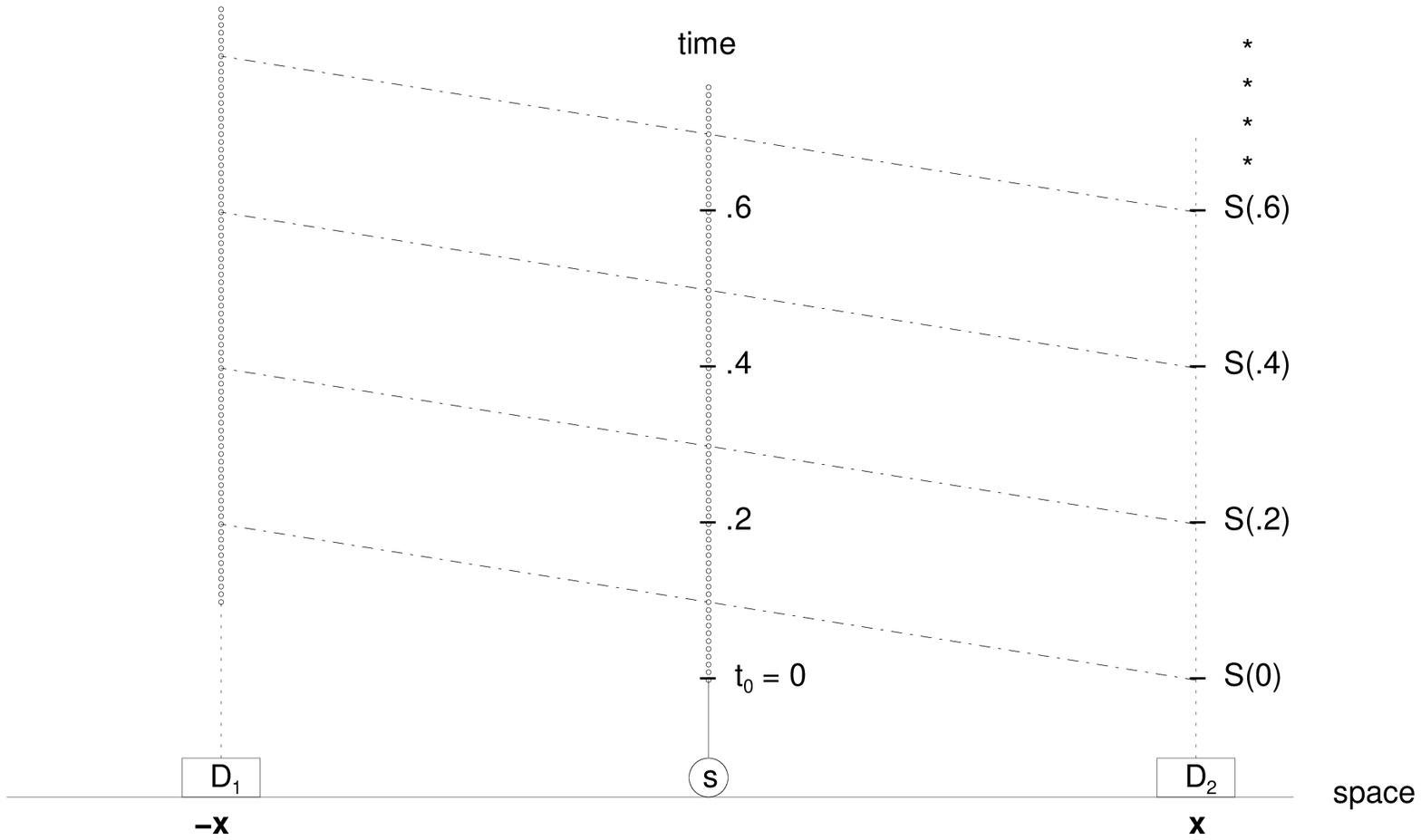}}\\
figure 19
\end{center}
Assuming the selection timing can be realized under lab conditions, the accompanying
spatial constraint remains considerable -  a detector separation of  $ 1.6 \times 10^8 $ m , about twice the distance to the moon.

\section{Conclusions}

In light of the in every case discouraging experimental constraints imposed by the requirements of the Bell inequality it is worth clarifying what is at issue and what precisely may EPR experimental results resolve; just what is at stake? 

\noindent The constraints originate from the EPR locality condition. It should be understood however that the question of physical locality or even that of determinism was secondary in the thinking and motivation behind EPR's critique of the orthodox view, used primarily as a means to show the incompleteness of quantum mechanics \cite{16,30, 34}. At bottom was a philosophic issue; that of the nature of reality: whether reality can be said to exist apart from the perception of it, EPR taking the affirmative view. In the opposition, the orthodox view \cite{34}. These philosophic leanings have been called respectively realistic and idealistic ( or logically positivistic) \cite{ 16, 33, 40}. We expand the previous schematic to include this background dimension
\begin{center}
{\bf pre-EPR }
\end{center}
 \begin{description}
\item{i)} {\sl realist argument }: realist world view (real ontology) \\ 
$\Rightarrow $ existence of real-unmeasured quantities, QM incomplete
\item{ii)} {\sl idealist argument }: idealist world view (ideal ontology) \\
$ \Rightarrow $  non-existence of unobserved quantities, QM complete
\item{{\sl result}}: No means to verify of dispel either world view; stalemate.
\end{description}
\begin{center}
{\bf post-EPR }
\end{center}
\begin{description}
\item{i)}{\sl realist argument }:
\begin{description}
\item{a)} Locality  \\ $ \Rightarrow $ existence of real-unmeasured quantities
\item{b)} Locality + counterfactual definiteness \\ 
$ \Rightarrow $ existence of simultaneous real-unmeasured quantities \\ $ \Rightarrow $ QM incomplete \& BI ( $ \Rightarrow $ QM incorrect )
\end{description}
\item{ii)} {\sl idealist argument }:
\begin{description}
\item{a)} Non-existence of unmeasured quantities \\ $ \Rightarrow $ non-locality
\end{description}
\item{{\sl result }}: Bell inequality as necessary condition for local realism
\end{description}
And so should faithful EPR experimental data verify BI, then the predictions of quantum mechanics will be shown incompatible with the results from certain detector setting combinations. On the other hand, should the data violate the BI, then it is rather local determinism that is found incompatable. Such an outcome might be understood as resulting from a non-local determinism, e.g., that of Bohm's \cite{45}, or a local indeterminism or indeed a nonlocal indeterminism, according to one's philosophical leanings... The initial question, that of the reality of unobserved phenomena, is in both cases left unanswered.

\noindent Non-locality aside, whether a violation of the BI would mean that we live in a counterfactually definite world, or equivalently, would leave room for the existence of hidden variables is yet another question, and one best answered in the affirmative by demonstration of the variables themselves in action, e.g., by way of position and momentum predictions for a Young's double-slit type experiment. The negative proof however would not be as simple, as empirical proofs against logically consistent possibilities are necessarily inductive, requiring that each out of an in principle infinite number of possibilities be eliminated. Small wonder then that in over half a century the hidden variable question has not gone away despite the best efforts \cite{4, 5, 6, 9, 11, 17, 18, 19, 36}. One will do well to observe the advice of philosopher Wittgenstein that ``as there is only a {\em logical} necessity, so there is only a {\em logical} impossibility'' ( such as a square-circle), and consider whether a proof against the existence of hidden variables could be more convincing than the many standing proofs against...  against e.g. the reality of paranormal phenomena\footnote{ See for example http://skepdic.com/   }, which is a different question from whether one believes there to be such phenomena. For logical necessity, Wittgenstein continues, is tautological \cite{13}. There is a nice quote in this regard from Pauli in a letter to Born
\begin{quotation}
\small

\noindent One should no more wrack one's brain about the problem of whether something one cannot know anything about exists all the same, than about the ancient question of how many angels are able to sit on the point of a needle. But it seems to me that Einstein's questions are ultimately always of this kind.\ \ldots

\end{quotation}
The tautology might run: Physical variables shall not be hidden. Hidden variables therefore do not exist \footnote{ On the other hand Einstein enjoyed only the most affectionate respect among contemporaries, as he does among today's scientists - a respect however that sometimes has amusing effects... By all accounts it was only in deference to Einstein (an otherwise brilliant man!) that his contemporaries were persuaded to consider the Lorentz locality restriction - {\em as long as one does not take it too far }. The restriction is generally regarded in orthodox circles as a mild embarrassment, one to be duly recognized then quickly dispatched, often by means of a few slighting remarks on the views of their realist opponents ( "conspiratorial", ``paranoid'', ``highly artificial'', and so forth.) that tend to appeal to classical intuitions \cite{33, 30, 5, 38}.}. It is an irony of history that the reticent philosophy of Wittgenstein has come to be associated with that of the Copenhagen view.

\noindent The question whether a tree falling in the forest makes a sound if no one is there to hear it has positivistic meaning only in the context of a world view that allows for causal irrelevance; likewise whether the moon is really there if no one is looking \cite{19}. To illustrate the difficulty for the scientist, the first of these might also be rendered " If a tree falls in the forest and you're not able to verify it, can you verify it?" And what then might one ask of the {\em existence} of hidden variables? Inquiries such as these are outside the domain of emperical resolution, and as such best left to philosophy whose business is precisely their elucidation and analysis\footnote{ Opposing sides on the philosophic question, as previously mentioned, are often labeled idealist and realist; the debate in its present form has endured two centuries with no resolution in sight. For a historical account see e.g. Bertrand Russell's { \sl The Problems of Philosophy }, New York: Oxford University Press (1959); http://www.ditext.com/russell/russell.html }. They do not belong to science.

\subsection{ Minimal-mathematics derivations}

Lastly, there is a type of BI derivation free of formal mathematics that seems to have a popular appeal; it can be found e.g. on the internet \cite{11,20}. The inquiring student should keep in mind that only two EPR measurements may be taken per particle-pair, i.e., per person, per playing card, per pet, etc., whatever the particle-pair analogue in a given derivation. This is so because the set of three measurements taken over the group represented in the BI is actually a set of three measurements taken respectively over three mutually exclusive subgroups.

\noindent  Consider for example a derivation in which the quantities to be measured are certain physical traits of individual humans of a group. For example, their 1) height, with possible values: tall(t) and short(s),  2) eye color, with possible values: blue(b), green(g), red( r), etc., 3) gender, with possible values: male(m), female(f), etc.. The measurements could take the form of a questionnaire. One might then be challenged\cite{20} to find any group for which the inequality
\bea
                            N( t, not-b) + N( b, not-m)  & \geq &  N( t, not-m)  \nonumber
\eea
does not hold. But alas, there are no such groups. In light of the preceding the clever student might now ask whether with prior knowledge of which question-pair would be put to which of the three sub-groups {\em in an actual survey } might it not be possible to shuffle the members around so as to out-maneuver the inequality, to exploit the given loophole. An excellent example of this possibility is revealed in another derivation of this type\cite{44} in itself worth careful consideration. The author adds "Bell's inequalities... give conditions on when a set of marginal probability distributions could have been derived from a single joint distribution " apparently oblivious to EPR experimental limitations. To repeat, now in this language, the marginal probability distributions may indeed be considered derived from an underlying single joint distribution, but only on the condition that its constituent sample-space measurements are made randomly, that they remain mutually independent of the joint distribution. And it is in turn the above Lorentz constraints derived from the above self-consistent semifactual that insures this independence. What the Bell Inequality proves from ``EPR'' experiments to date beyond a doubt is that local hidden variables, should they exist, are, if anything, clever - no less clever than the microscopic phenomena they describe is peculiar.

\end{document}